\documentclass{aa}

\usepackage{graphicx}
\usepackage{natbib}
\usepackage{txfonts}
\usepackage[colorlinks=true,allcolors=blue]{hyperref}

\newcommand{\bz}{$\langle B_\mathrm{z} \rangle$}
\newcommand{\bm}{$\langle B \rangle$}
\newcommand{\teff}{$T_{\rm eff}$}
\newcommand{\lgg}{$\log g$}
\newcommand{\vs}{$\varv_{\rm e}\sin i$}
\newcommand{\kms}{km\,s$^{-1}$}

\newcommand{\orcidlink}[1]{\protect\href{https://orcid.org/#1}{\protect\includegraphics[width=8pt]{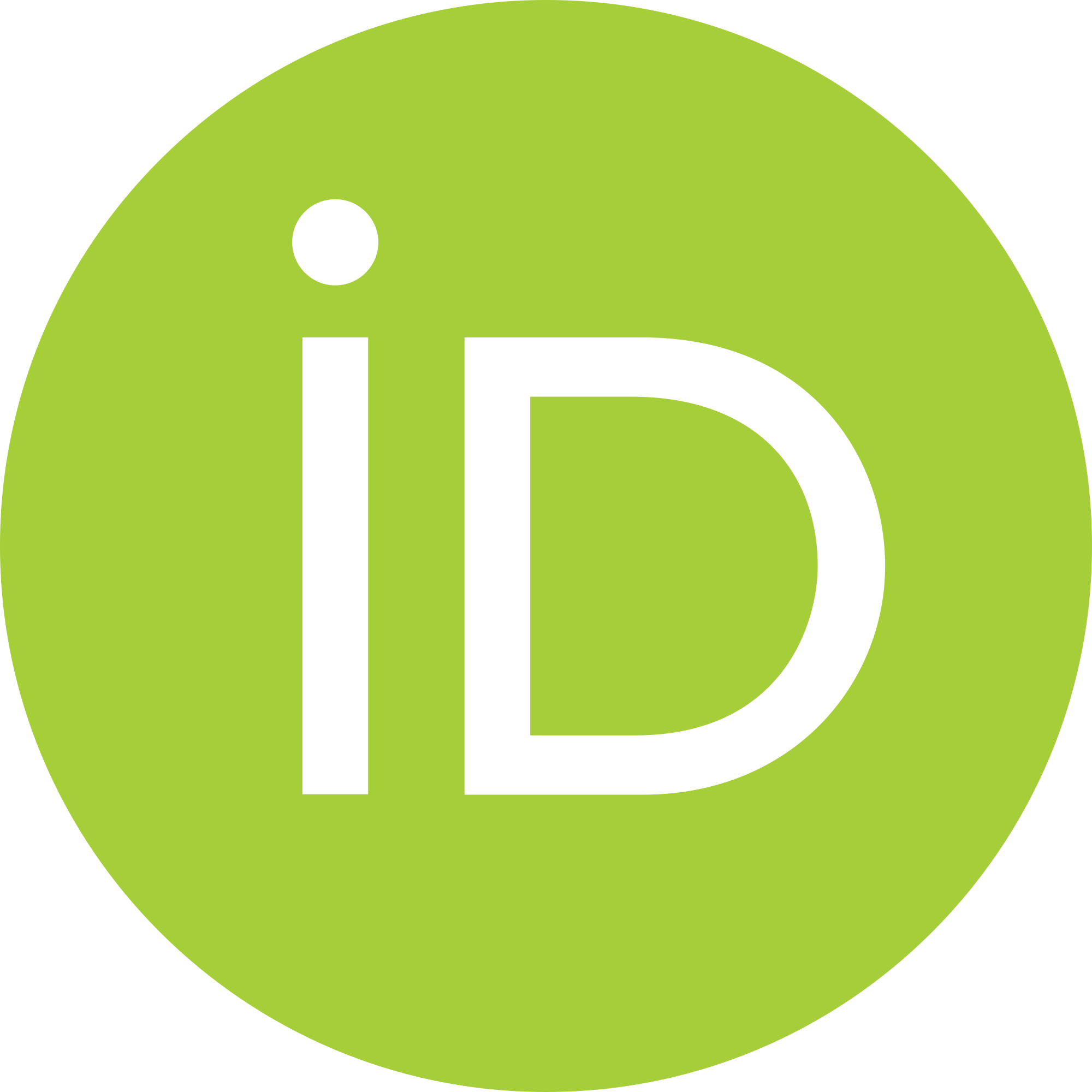}}}

\newcommand{\fifps}[2]{\centering\resizebox{#1}{!}{\includegraphics{#2}}}
\newcommand{\fihps}[2]{\centering\resizebox{!}{#1}{\includegraphics{#2}}}

\newcommand{\beq}{\begin{equation}}
\newcommand{\eeq}{\end{equation}}

\begin{document}

\title{Zeeman--Doppler imaging of early-type stars}
\subtitle{Testing Stokes parameter inversions with theoretical magnetohydrodynamic models}

\author{%
O.~Kochukhov\inst{1}\corrauth{oleg.kochukhov@physics.uu.se}\orcidlink{0000-0003-3061-4591}
\and A.~Al~Kanale\inst{1}
\and J.~Braithwaite\inst{2}
}

\institute{
Department of Physics and Astronomy, Uppsala University, Box 524, SE-75120 Uppsala, Sweden
\and
Argelander-Institut f\"ur Astronomie, Universit\"at Bonn, Auf dem H\"ugel 71, D-53121 Bonn, Germany
}

\date{Received 22 September 2026 / Accepted 2 November 2026}

\titlerunning{Zeeman--Doppler imaging of early-type stars}
\authorrunning{O.~Kochukhov et al.}

\abstract
{
Zeeman--Doppler imaging (ZDI) is widely used to reconstruct magnetic fields in early-type stars from time-series spectropolarimetric observations, yet its accuracy has not been systematically validated against physically self-consistent, complex magnetic field models.
}
{
We aim to assess the reliability and limitations of modern ZDI techniques applied to fossil fields of early-type stars.
}
{
We generated synthetic spectropolarimetric data from three-dimensional magnetohydrodynamic (MHD) simulations of fossil magnetic fields. The resulting Stokes profiles -- individual lines and least-squares deconvolved (LSD) profiles -- were inverted with contemporary ZDI codes using Stokes $IQUV$ and $IV$ data. We compared the reconstructed maps with the input MHD surface fields and quantified deviations in field strength, vector orientation, and harmonic energy decomposition.
}
{
For Stokes $IQUV$ inversions, ZDI recovers field vector components and strength with better than 10\% accuracy, while Stokes $IV$ inversions degrade to 15--20\% for the most complex geometries. The field inclination is recovered to within 10\degr, with errors increasing to 15--20\degr\ in the same cases. Full Stokes inversions reproduce poloidal and toroidal energies to within a few percentage points. LSD-based inversions yield accuracy comparable to, and sometimes better than, that achieved with ZDI using individual \ion{Fe}{ii} lines. Including linear polarisation offers a clear advantage for complex geometries. Imposing a current-free poloidal constraint, which the input field satisfies, only helps where the unconstrained Stokes $IV$ inversion performs poorly. Cross-talk between magnetic and abundance reconstructions is weak, with spurious inhomogeneities at or below 0.1 dex on average.
}
{
Modern ZDI robustly reconstructs fossil magnetic fields in early-type stars, including their smaller-scale non-dipolar features, particularly when full Stokes information is available. Stokes $IV$-only inversions, however, can introduce measurable artefacts for complex topologies. ZDI reliably recovers the harmonic content of global fields, successfully disentangling poloidal and toroidal contributions. These results establish quantitative benchmarks for modern ZDI applied to early-type magnetic stars.
}


\maketitle
\nolinenumbers
 
\section{Introduction}
\label{sec:intro}

Zeeman--Doppler imaging \citep[ZDI;][]{kochukhov:2016} is a tomographic technique that reconstructs the surface magnetic field topology of a star from a time series of spectropolarimetric observations. The method exploits the Zeeman effect, which causes spectral lines to split and become polarised in a magnetic field. As a star rotates, magnetic surface features move across the visible disk, and the resulting Doppler shifts encode the spatial distribution of the field into the rotationally modulated line profiles. By analysing the time-dependent polarised signatures in these profiles, typically circular polarisation (Stokes $V$), and sometimes linear polarisation (Stokes $Q$ and $U$), ZDI recovers the strength and orientation of the large-scale field; in doing so it distinguishes poloidal and toroidal components and quantifies the relative contributions of dipolar, quadrupolar, and higher-order modes. The technique is particularly well suited to magnetic early-type stars such as the chemically peculiar Ap and Bp stars, which host strong, organised, and stable fossil fields \citep{donati:2009}. It also enables the joint mapping of magnetic fields and surface chemical abundance spots, which trace how the field channels element diffusion in stellar atmospheres \citep{michaud:2015}. The reconstructed field topologies constrain the origin and long-term evolution of magnetism in intermediate- and high-mass stars.

The concept of ZDI goes back to the pioneering studies of \citet{piskunov:1983,piskunov:1984} and \citet{piskunov:1985}, who formulated stellar magnetic mapping as an ill-posed inverse problem and developed methods to reconstruct surface magnetic field distributions from spectropolarimetric observations. These techniques were soon used to model high-resolution circular polarisation measurements of the Ap star $\alpha^2$~CVn \citep{glagolevskii:1985a}. Groundbreaking as it was, this early work relied on analytical approximations for local Stokes profiles and was restricted to simple dipolar magnetic field geometries. Progress was further hindered by the lack of spectropolarimetric observations of sufficient quality and completeness for more sophisticated reconstructions.

Magnetic mapping was subsequently extended to the more complex dynamo-generated magnetic fields of late-type active stars \citep{semel:1989,brown:1991,donati:1997a}. This implementation of ZDI still relied on analytical representations of local spectral line profiles but described the stellar magnetic field via discrete surface maps of the three vector field components, which were constrained by the maximum-entropy regularisation but otherwise capable of accommodating magnetic structures of arbitrary complexity. A major advance was the incorporation of the least-squares deconvolution technique \citep[LSD;][]{donati:1997}, which combined the polarisation signatures of many spectral lines into a single mean profile with a high signal-to-noise ratio (S/N), replacing the noisy measurements from individual lines by a time series of high-quality mean polarisation profiles. Together, these developments made ZDI a practical tool for cool-star magnetic tomography studies and enabled its first successful applications to a variety of active late-type stars \citep[e.g.][]{donati:1999,donati:1999b,donati:2003}.

For magnetic Ap stars, the next major advance in ZDI methodology was introduced by \citet{piskunov:2002a}, who incorporated detailed numerical calculations of Stokes parameter spectra -- based on polarised radiative transfer in realistic stellar model atmospheres -- into the inversion procedure. This enabled a self-consistent reconstruction of stellar magnetic fields of arbitrary complexity simultaneously with maps of chemical abundance inhomogeneities. A comprehensive suite of numerical tests of this new ZDI implementation \citep{kochukhov:2002c} demonstrated that the method reliably recovers low-order multipolar magnetic field geometries together with abundance distributions consisting of one or several large-scale spots, provided that observations in all four Stokes parameters are available. As in contemporary ZDI studies of cool stars, the magnetic field was represented by independent surface maps of the three vector field components and constrained only through Tikhonov regularisation. However, similarly to the findings of \citet{donati:1997a}, these studies highlighted the difficulty of recovering smooth, large-scale magnetic structures from the less informative Stokes $V$ observations alone. This limitation was particularly relevant in practice, since observational datasets rarely included linear polarisation measurements in Stokes $Q$ and $U$. To mitigate this problem, \citet{piskunov:2002a} introduced, and \citet{kochukhov:2002c} validated, a multipolar regularisation scheme that biased the reconstructed field towards the optimal non-axisymmetric superposition of low-order poloidal dipolar and quadrupolar components \citep{landolfi:1998}. Subsequent ZDI studies of Ap stars employed two complementary versions of the methodology: essentially assumption-free, discrete magnetic field reconstruction for inversions based on full Stokes $IQUV$ observations \citep{kochukhov:2004d,kochukhov:2010,silvester:2014} and inversions incorporating multipolar regularisation when only Stokes $IV$ spectra were available \citep{kochukhov:2002b,luftinger:2010}.

A further development in ZDI methodology was the general spherical harmonic description of the stellar magnetic field introduced by \citet{donati:2006b}. This description automatically satisfies the solenoidal condition and provides a natural framework for quantifying the poloidal and toroidal components of the field at different angular scales, offering a more compact and physically motivated parameterisation of global magnetic field topologies than discrete maps of independent vector field components. \citet{kochukhov:2014} subsequently implemented an equivalent parameterisation based on real spherical harmonics in Ap-star ZDI; the authors also developed a method for incorporating LSD profiles without sacrificing the physical realism of the polarised radiative transfer calculations, which substantially reduced the observational requirements of the technique. This updated ZDI formulation has since been widely adopted in studies of magnetic early-type stars, both for inversions based on individual spectral lines \citep[e.g.][]{silvester:2015,silvester:2017,rusomarov:2015,rusomarov:2016,rusomarov:2018} and for analyses employing LSD profiles, either in Stokes $IV$ \citep[e.g.][]{kochukhov:2022,kochukhov:2023a,kochukhov:2025} or in all four Stokes parameters \citep[e.g.][]{kochukhov:2017a,kochukhov:2019}.

Hot-star ZDI has evolved substantially since the studies of \citet{piskunov:2002a} and \citet{kochukhov:2002c}, to the extent that some aspects of the numerical experiments reported in the latter work are now outdated. The problem of chemical spot mapping was revisited by \citet{kochukhov:2017}, who presented an updated suite of tests quantifying the systematic biases in abundance inversions: those arising from the neglect of feedback between local chemical anomalies and the atmospheric structure, and those introduced when magnetic fields are not accounted for in abundance mapping. Equivalent modern tests for magnetic field inversions remain unavailable, a gap that is particularly relevant given the increasing sophistication of modern ZDI implementations and the improved quality of spectropolarimetric observations. The main outstanding issues concern (i) how accurately current methods recover distorted dipolar fields combined with smaller-scale magnetic structures, using either full Stokes datasets or only Stokes $IV$ spectra; (ii) whether the harmonic ZDI formulation reliably reproduces the relative contributions of poloidal and toroidal field components across different angular scales; and (iii) whether the transition from modelling individual spectral lines to using LSD profiles introduces additional uncertainties or systematic biases that may remain unrecognised within the inversion procedure.

Fortunately, the rapid progress in the observational characterisation of hot-star magnetic fields has been matched by equally substantial theoretical developments. \citet{braithwaite:2004}, \citet{braithwaite:2006}, and \citet{braithwaite:2008} presented a series of three-dimensional magnetohydrodynamic (MHD) simulations of the relaxation (in stably stratified stellar envelopes) of arbitrary initial fields into stable equilibria, which can then in principle be observed as fossil fields. These calculations demonstrated that the long-term stability of fossil-field configurations requires the coexistence of poloidal and toroidal field components within stellar interiors, and that this relaxation naturally gives rise to a wide range of complex surface magnetic topologies. These physically self-consistent, first-principles MHD models therefore provide an ideal starting point for numerical validation tests of ZDI aimed at addressing the questions outlined above. Our aim is to synthesise spectropolarimetric observations from surface magnetic field maps taken from these simulations and to invert them with current ZDI methodologies, so that a comparison of the reconstructed and input maps allows us to assess the accuracy and precision of the inversions under different conditions and to reveal potential systematic biases. The benchmarking of ZDI performance against theoretically motivated MHD models is well established in investigations of cool-star, dynamo-generated magnetic fields \citep{yadav:2015,lehmann:2019,hackman:2024}, but it has not previously been extended to the fossil magnetic fields of hot stars.

The paper is structured as follows. Section~\ref{sec:method} presents the theoretical foundations, algorithmic implementation, and key numerical details of ZDI for several different inversion codes. Section~\ref{sec:experiments} introduces the MHD fossil-field models used for the numerical validation of ZDI and describes the corresponding inversion experiments based on both individual spectral lines and LSD profiles. Section~\ref{sec:conclusions} summarises the main results and discusses their implications.

\section{Zeeman--Doppler imaging methodology}
\label{sec:method}

\subsection{Polarised spectrum synthesis}
\label{sec:synthesis}

Detailed spectrum synthesis in all four Stokes parameters, based on realistic one-dimensional stellar model atmospheres, is a fundamental component of our ZDI methodology. Here we briefly summarise the underlying physical assumptions and key algorithmic choices adopted in these calculations. The numerical implementation and radiative transfer formalism are described in detail by \citet{piskunov:2002a} and \citet{kochukhov:2014}.

Theoretical spectra are computed using model atmospheres from the {\sc ATLAS} grids \citep{kurucz:2005} or dedicated {\sc LLmodels} atmospheres \citep{shulyak:2004} for hot stars, and from the {\sc MARCS} grid \citep{gustafsson:2008} for cool stars. Continuum opacities, line-centre opacities, and the corresponding Gaussian and Lorentzian broadening parameters are calculated from the thermodynamic structure of the adopted model atmosphere for a specified set of chemical abundances, which may be depth-dependent. Depending on the application, we either adopt a single model atmosphere for the entire stellar surface or interpolate within a grid of atmospheres computed for different temperatures or abundances.

The resulting line opacity is multiplied by the absorption and anomalous-dispersion profiles obtained by summing over the individual Zeeman components of all spectral lines contributing to a given wavelength interval. For hot stars, the local abundance variations of the mapped chemical species are treated by scaling the corresponding line absorption profiles. In this approximation, the effects of abundance variations on the atmospheric structure, ionisation balance, and continuum opacity are neglected. For cool stars, we employ the self-consistent molecular equilibrium solver described by \citet{piskunov:2017}, which enables realistic modelling of the effects of starspots on both line and continuum spectra. This self-consistent approach can also be applied to abundance mapping in hot stars, treating one chemical element at a time, provided that the adopted model atmosphere grid incorporates the corresponding abundance variations. In practice, however, this refinement is generally unnecessary for elements other than helium \citep{kochukhov:2012,kochukhov:2017}.

The spectral line data, together with the $J$ quantum numbers and Land\'e factors of the atomic levels needed to compute Zeeman splitting patterns, are obtained from the Vienna Atomic Line Database \citep[VALD;][]{ryabchikova:2015}. Our calculations can account for partial Paschen--Back splitting \citep{kochukhov:2008a} and departures from local thermodynamic equilibrium for selected spectral lines. These effects are incorporated through externally supplied tables of magnetic splitting patterns and departure coefficients, respectively \citep[e.g.][]{oksala:2015}.

Once the line and continuum opacities have been determined, we construct the full polarised absorption matrix for the range of limb angles and corresponding magnetic field vector orientations with respect to the line of sight that a given surface element assumes as the star rotates. The polarised radiative transfer equation is then solved using the DELO-B\'ezier formal solver \citep{de-la-cruz-rodriguez:2013}. The emergent Stokes parameter spectra are calculated on an adaptive wavelength grid and interpolated onto a fine equidistant grid, where they are convolved with radial-tangential macroturbulent and Gaussian instrumental broadening profiles. In these calculations, the magnetic field strength and orientation are assumed to be depth-independent.

\subsection{Surface grid and disk integration}
\label{sec:grid}

To calculate the emergent Stokes flux spectra of a rotating star with an arbitrary magnetic field topology, we discretise the stellar surface into 1000--4000 elements following the scheme of \citet{piskunov:2002a}. This grid is constructed with a latitude-dependent number of longitudinal divisions, so that all elements have approximately equal areas. For each rotational phase, the magnetic field vector in every surface element is transformed from the stellar reference frame to the observer's coordinate system. Polarised radiative transfer calculations are then carried out as described above. The resulting local Stokes spectra are Doppler-shifted according to the projected rotational velocity of each element. By default, rigid rotation is assumed, although parametric differential rotation can be incorporated through the surface velocity field. Line and continuum spectra are then summed over the visible stellar disk with weights proportional to the projected areas of the corresponding surface elements. No parametric limb-darkening law is employed, as the angular dependence of the emergent Stokes spectra is obtained directly from the radiative transfer calculations.

To maintain the accuracy of the disk integration despite the relatively coarse surface discretisation, particularly for projected rotational velocities exceeding $\sim$100~\kms, each element is further subdivided into many sub-elements. This procedure accounts for the variation of rotational Doppler shifts across an element by assigning the same local Stokes spectrum to all sub-elements, applying the appropriate Doppler shift to each, and summing their contributions. The same subdivision scheme is also used to evaluate accurate projected areas for elements at the limb.

Our ZDI codes additionally support temporal averaging over the finite exposure time of an observation. In this mode, each rotational phase is subdivided into three to five closely spaced sub-phases spanning the integration interval centred on the nominal phase. Synthetic spectra are computed for each sub-phase and combined using trapezoidal integration. This procedure accounts for rotational phase smearing when the exposure time is a non-negligible fraction of the stellar rotation period.

Some of our ZDI implementations include extensions to the basic forward-modelling framework presented here. These comprise the treatment of non-radial pulsational velocity fields in addition to rotational broadening, as well as the effects of rapid rotation, including rotational flattening, gravity darkening, and the resulting changes of element areas and surface normals. Extensions of ZDI to double-lined eclipsing binary stars with Roche-lobe geometries have also been developed \citep{kochukhov:2019a,hahlin:2021}, but these have so far been applied only to late-type active binaries and are therefore not considered here.

\subsection{Least-squares deconvolution}
\label{sec:lsd}

Least-squares deconvolution \citep{donati:1997,kochukhov:2010a} is a widely used multi-line technique that optimally co-adds weak polarisation signatures from thousands of spectral lines, substantially increasing the S/N of the polarisation signal and making it suitable for detailed modelling. Even for strong-field early-type stars, LSD greatly simplifies the analysis of Stokes $V$ and, in particular, Stokes $Q$ and $U$ spectra compared to analyses of individual spectral lines \citep{wade:2000b,silvester:2012}.

The LSD method assumes that spectral line profiles are self-similar, which is strictly valid only in the weak-field and weak-line limits \citep{kochukhov:2010a}. Nevertheless, LSD profiles of strong-field stars can be modelled without compromising generality or physical realism, provided that LSD is applied consistently to both the observed (or simulated) Stokes parameter spectra and the model synthetic spectra spanning the full wavelength range over which the LSD profiles are computed \citep{kochukhov:2014,kochukhov:2026}. This approach is adopted here and tested in the context of ZDI in Sect.~\ref{sec:test_lsd}. We first calculate synthetic full-Stokes spectra over a broad wavelength range including all relevant spectral lines for the given stellar atmospheric parameters. This is achieved by synthesising local Stokes spectra with the {\sc Synmast} code \citep{kochukhov:2010a}, followed by disk integration using a custom code written in the Interactive Data Language (IDL). Random noise is then added to the resulting flux spectra to simulate observational uncertainties. Finally, LSD profiles are extracted from these simulated observations using the {\sc iLSD} code \citep{kochukhov:2010a}\footnote{\url{https://github.com/okochukhov/iLSD}} and the line mask of the element of interest. In our approach, blending lines of chemical elements other than the element of interest are included in the synthetic spectra and are accounted for with the multi-profile version of LSD implemented in this code. These lines are represented by a single additional mean profile, derived simultaneously with the element-specific one. The resulting element-specific LSD profiles are the input to the inversion code.

On the inversion side, we again use {\sc Synmast} to calculate the local specific-intensity spectra for all four Stokes parameters over the same wavelength range and with the same line list as for the simulated observations. LSD is subsequently applied in the same way to these local Stokes spectra, and the resulting mean line profiles are tabulated as functions of element abundance, limb angle, magnetic field strength, and the angle between the field vector and the line of sight. This calculation can be made fully consistent with the atmospheric structure by using a separate model atmosphere for each abundance value. The mean local continuum intensity required for spectrum synthesis is likewise tabulated as a function of element abundance and limb angle. The construction of the local LSD profile table is time-consuming, but it is performed only once for a given set of stellar atmospheric parameters and can then be reused for multiple inversions. During the inversion, the code linearly interpolates within these precomputed tables whenever the Stokes parameters are synthesised.

\subsection{Harmonic representation of the stellar magnetic field}
\label{sec:harmonic}

Following \citet{donati:2006b} and \citet{kochukhov:2014}, we represent the stellar magnetic field using a general spherical harmonic expansion. This approach allows a field structure of arbitrary complexity to be described while guaranteeing that the field remains solenoidal, with zero net magnetic flux through the stellar surface. The harmonic parameterisation also enables a convenient quantitative assessment of the contributions of different harmonic components associated with different spatial scales.

The general harmonic expansion implemented in our ZDI codes describes the field as the sum of poloidal and toroidal contributions,
\beq
\vec{B} (\theta, \varphi) = \vec{B}^{\rm pol} (\theta, \varphi) + \vec{B}^{\rm tor} (\theta, \varphi),
\eeq
with the poloidal field given by
\begin{align}
B^{\rm pol}_r (\theta, \varphi) = & -\sum_{\ell=1}^{\ell_{\rm max}} \sum_{m=-\ell}^{\ell} \alpha_{\ell,m} Y_{\ell,m} (\theta, \varphi) \nonumber \\
B^{\rm pol}_\theta (\theta, \varphi) = & -\sum_{\ell=1}^{\ell_{\rm max}} \sum_{m=-\ell}^{\ell} \beta_{\ell,m} Z_{\ell,m} (\theta, \varphi) \\
B^{\rm pol}_\varphi (\theta, \varphi) = & -\sum_{\ell=1}^{\ell_{\rm max}} \sum_{m=-\ell}^{\ell} \beta_{\ell,m} X_{\ell,m} (\theta, \varphi), \nonumber
\end{align}
and toroidal components described with
\begin{align}
B^{\rm tor}_r (\theta, \varphi) = & \phantom{-} 0 \nonumber \\
B^{\rm tor}_\theta (\theta, \varphi) = & -\sum_{\ell=1}^{\ell_{\rm max}} \sum_{m=-\ell}^{\ell} \gamma_{\ell,m} X_{\ell,m} (\theta, \varphi) \\
B^{\rm tor}_\varphi (\theta, \varphi) = & \phantom{-}\sum_{\ell=1}^{\ell_{\rm max}} \sum_{m=-\ell}^{\ell} \gamma_{\ell,m} Z_{\ell,m} (\theta, \varphi). \nonumber
\end{align}
The expansion is defined on the stellar surface, with no explicit radial dependence assumed for the magnetic field geometry.

The quantities $Y_{\ell,m} (\theta, \varphi)$, $Z_{\ell,m} (\theta, \varphi)$, and $X_{\ell,m} (\theta, \varphi)$ are defined in terms of real spherical harmonics and their derivatives for angular degree $\ell$ and order $m$,
\beq
Y_{\ell,m} (\theta, \varphi) = -C_{\ell,m} P_{\ell,|m|} (\cos\theta) K_m (\varphi),
\eeq
\beq
Z_{\ell,m} (\theta, \varphi) = \frac{C_{\ell,m}}{\ell+1} \frac{\partial P_{\ell,|m|} (\cos\theta)}{\partial \theta} K_m (\varphi),
\eeq
\beq
X_{\ell,m} (\theta, \varphi) = \frac{C_{\ell,m}}{\ell+1} \frac{P_{\ell,|m|} (\cos\theta)}{\sin \theta} |m| M_{m} (\varphi),
\eeq
with the normalisation factor
\beq
C_{\ell,m} = \sqrt{\frac{2\ell+1}{4\pi}\frac{(\ell-|m|)!}{(\ell+|m|)!}},\eeq
and
\beq
K_m (\varphi) = \left\{  
\begin{array}{rl}
\cos(|m|\varphi), & m \ge 0 \\
\sin(|m|\varphi), & m < 0 \\
\end{array}\right.,
\eeq
\beq
M_m (\varphi) = \left\{  
\begin{array}{rl}
-\sin(|m|\varphi), & m \ge 0 \\
\cos(|m|\varphi), & m < 0 \\
\end{array}\right..
\eeq
In these expressions, $\theta$ and $\varphi$ denote the colatitude and longitude on the stellar surface, respectively, and $P_{\ell,m} (\cos \theta)$ is the associated Legendre polynomial. These angles are measured with respect to the stellar rotational axis, but neither the separation of the field into poloidal and toroidal parts nor its energy spectrum in $\ell$ depends on this choice.

The free parameters of this harmonic expansion are the coefficients $\alpha_{\ell,m}$, $\beta_{\ell,m}$, and $\gamma_{\ell,m}$, which characterise the radial poloidal, horizontal poloidal, and horizontal toroidal magnetic field components, respectively. The expansion is carried out up to a sufficiently high $\ell_{\rm max}$ in order to reproduce the level of detail present in the observed Stokes profiles. Here we adopted $\ell_{\rm max}=10$, which corresponds to 360 independent harmonic coefficients, in line with typical choices in ZDI studies of early-type stars.

As discussed by \citet{kochukhov:2016a}, not all families of spherical harmonic coefficients are necessarily required to describe stellar magnetic fields in all cases. Historical analyses of magnetic topologies of early-type stars based on fitting integral magnetic observables \citep[e.g.][]{bagnulo:2000,bagnulo:2002} were limited to $\ell_{\rm max}=2$, neglected toroidal field components ($\gamma_{\ell,m}=0$), and adopted linked horizontal and radial poloidal components ($\alpha_{\ell,m}=\beta_{\ell,m}$). Together, the latter two restrictions correspond to a surface field that matches a current-free exterior extending to infinity. A more realistic representation of the stellar magnetosphere is provided by the potential source surface field extrapolation \citep{hussain:2002,jardine:2013}, which keeps the field current-free only out to a finite radius, beyond which the outflowing stellar wind is assumed to drag it into a purely radial configuration. This outer boundary condition unlinks the radial and horizontal poloidal components, yielding $\alpha_{\ell,m}\ne\beta_{\ell,m}$. Here we test ZDI based on the most general harmonic expansion involving all three families of coefficients, and also assess the impact of restricting the poloidal parameterisation to $\alpha_{\ell,m}=\beta_{\ell,m}$.

\subsection{Regularised inverse problem solution}
\label{sec:inverse}

The recovery of the surface structure of a star from a time series of Stokes parameter spectra is an ill-posed inverse problem. Owing to the finite S/N of the observations, gaps in the rotational phase coverage, and the limited information content of incomplete sets of polarisation profiles, an infinite number of surface distributions reproduce the data within the observational uncertainties, many of them exhibiting large-amplitude small-scale structure. A unique and physically meaningful solution can therefore be identified only if the inversion is stabilised by additional a priori constraints. Following \citet{piskunov:2002a}, we obtain such a solution by minimising the total functional
\begin{align}
\Psi &=\sum_X C_X \sum_{\varphi,\lambda} \left[X_{\varphi,\lambda}^{\rm obs} -
X_{\varphi,\lambda}^{\rm syn}\left(\alpha_{\ell,m}, \beta_{\ell,m}, \gamma_{\ell,m}, \varepsilon_{i}^{(n)}\right)\right]^2 / \delta X_{\varphi,\lambda}^2
\nonumber\\
&+ C_{\rm z} \sum_{\varphi} \left[\langle B_{\rm z} \rangle_{\varphi}^{\rm obs} - \langle B_{\rm z} \rangle_{\varphi}^{\rm syn} \left(\alpha_{\ell,m}, \beta_{\ell,m}, \gamma_{\ell,m}\right) \right]^2 / \delta\langle B_{\rm z} \rangle_{\varphi}^2 \label{eq:psi} \\
&+ \sum_n \Lambda_{\varepsilon} R_{\varepsilon}(\varepsilon_{i}^{(n)}) + \Lambda_B R_B \left(\alpha_{\ell,m}, \beta_{\ell,m}, \gamma_{\ell,m} \right) \nonumber
\end{align}
with respect to all free parameters of the model. These comprise the spherical harmonic coefficients $\alpha_{\ell,m}$, $\beta_{\ell,m}$, and $\gamma_{\ell,m}$ describing the magnetic field topology and the values $\varepsilon_{i}^{(n)}$ of scalar surface parameters, which correspond to logarithmic chemical abundances in the applications considered here, in each of the $N$ surface elements $i$ and for each of the chemical species $n$ included in the inversion, amounting to up to a few thousand unknowns in total. The global stellar parameters, such as the inclination of the rotational axis, the azimuth angle of that axis, the projected rotational velocity, and the rotational period, are established prior to the magnetic inversion and are held fixed. Some of these parameters can be constrained by running repeated inversions, as demonstrated by \citet{kochukhov:2002c}.

The first term of Eq.~(\ref{eq:psi}) is the usual $\chi^2$ measure of the mismatch between the observed and computed Stokes spectra. Its inner summation extends over all rotational phases $\varphi$ and all wavelength points $\lambda$ of the spectral intervals included in the analysis, with $\delta X_{\varphi,\lambda}$ denoting the observational uncertainties. The outer summation runs over the Stokes parameters available for a given dataset, i.e. $X=I, Q, U, V$ for full Stokes inversions and $X=I, V$ when only circular polarisation has been measured. For datasets with low-quality linear polarisation measurements, we also use a mode in which the Stokes $Q$ and $U$ profiles are replaced by the total linear polarisation $P_{\rm L}=\sqrt{Q^2+U^2}$. The coefficients $C_X$ adjust the relative contributions of individual Stokes parameters, which is necessary because an unweighted $\chi^2$ is dominated by Stokes $I$ at the expense of the magnetically sensitive polarisation profiles. For inversions based on LSD profiles (Sect.~\ref{sec:lsd}) the same expression is used, with $X_{\varphi,\lambda}$ representing mean Stokes profiles instead of individual line spectra.

The second term of Eq.~(\ref{eq:psi}) provides an optional constraint from measurements of the mean longitudinal magnetic field \bz. 
The corresponding model prediction is obtained by averaging the line-of-sight field component over the visible stellar hemisphere with weights given by the projected areas and local continuum intensities of the surface elements, and therefore depends on the harmonic coefficients alone. This constraint is useful when the \bz\ measurements are obtained from spectral lines of an element with a uniform surface distribution, such as hydrogen. Such data are not affected by chemical inhomogeneities and therefore provide complementary information that helps to anchor the global field geometry \citep{kochukhov:2011a}. In the numerical experiments presented below we relied on the simulated Stokes profile time series alone and set $C_{\rm z}=0$.

The final two terms of Eq.~(\ref{eq:psi}) are the regularisation functionals of the abundance distributions and of the magnetic field, scaled by the regularisation parameters $\Lambda_{\varepsilon}$ and $\Lambda_B$. For the abundance map of each chemical species we adopt the Tikhonov regularisation
\beq
R_{\varepsilon} = \sum_{k=1}^N \sum_{j=1}^M \left(\varepsilon_{k}^{(n)} - \varepsilon_{j}^{(n)}\right)^2 / d_{k,j}^2,
\eeq
where the outer summation is carried out over all $N$ surface elements, the inner one over the $M$ elements adjacent to the element $k$, and $d_{k,j}$ is the angular distance between the centres of the two elements. We typically use $M=4$. This functional is a discrete approximation of the surface integral of the squared gradient of the abundance distribution. Its minimisation selects the smoothest abundance map compatible with the observations. 

The magnetic field is regularised directly in the harmonic domain by means of the penalty function
\beq
R_B = \sum_{\ell,m} \ell^2 \left(w^r_{m} \alpha_{\ell,m}^2 + w^h_{\ell,m} \beta_{\ell,m}^2 + w^h_{\ell,m} \gamma_{\ell,m}^2 \right),
\eeq
where the weights
\beq
w^r_{m} =
\left\{
\begin{array}{ll}
1, & m = 0 \\
1/2, & m \ne 0 \\
\end{array}\right.
\eeq
and
\beq
w^h_{\ell,m} = \dfrac{\ell}{\ell + 1}
\left\{
\begin{array}{ll}
1, & m = 0 \\
1/2, & m \ne 0 \\
\end{array}\right.
\eeq
are proportional to the surface integrals of the squared harmonic functions introduced in Sect.~\ref{sec:harmonic}, $Y^2_{\ell,m}$ for the radial field component and $Z^2_{\ell,m} + X^2_{\ell,m}$ for the horizontal ones. These factors convert the squared harmonic coefficients into the corresponding contributions to the surface-integrated $B^2$, so that, apart from the explicit $\ell^2$ factor, harmonic modes of equal energy are penalised equally irrespective of their degree and order. Owing to the orthogonality of the harmonic functions, the sum $\sum_{\ell,m} ( w^r_{m} \alpha_{\ell,m}^2 + w^h_{\ell,m} \beta_{\ell,m}^2 + w^h_{\ell,m} \gamma_{\ell,m}^2 )$ is thus proportional to the magnetic energy density integrated over the stellar surface, with the three terms measuring the contributions of the radial poloidal, horizontal poloidal, and toroidal field components. 

The functional $R_B$ is therefore the magnetic energy weighted by $\ell^2$. Its minimisation suppresses the power of high-$\ell$ harmonic modes and biases the reconstruction towards the simplest field topology consistent with the observations. In contrast to the multipolar regularisation of \citet{piskunov:2002a}, it neither assumes the presence of a dominant low-order multipolar component nor couples the poloidal and toroidal field families. For the constrained parameterisation $\alpha_{\ell,m}=\beta_{\ell,m}$ discussed in Sect.~\ref{sec:harmonic}, the first two terms of $R_B$ merge into a single contribution with the weight $\ell^2 (w^r_{m} + w^h_{\ell,m})$. The two regularisation parameters, $\Lambda_{\varepsilon}$ and $\Lambda_B$, control the balance between the quality of the fit to the observations and the smoothness of the resulting surface maps. They are calibrated with a series of trial inversions, as described for abundance mapping by \citet{kochukhov:2017} and illustrated for $\Lambda_B$ in Sect.~\ref{sec:setup}.

The total functional is minimised with respect to all free parameters simultaneously with the help of the modified Marquardt--Levenberg algorithm described by \citet{piskunov:2002a}. In this scheme, corrections to the parameters are obtained by solving a linear system of equations involving the gradient of $\Psi$ and the curvature matrix, in which the second derivatives of the Stokes parameter $\chi^2$ terms are approximated by products of the first derivatives of the synthetic profiles, whereas the derivatives of the \bz\ term and of the regularisation functionals are evaluated using explicit analytical expressions. The derivatives of the local Stokes profiles with respect to the magnetic field vector and the local abundances are computed alongside the formal solution of the polarised radiative transfer equation and are converted into derivatives with respect to the harmonic coefficients by the chain rule. Each iteration starts with a calculation of the gradient and the curvature matrix. These are then kept fixed during several sub-iterations, in which the damping parameter controlling the interpolation between a gradient search and the linearised solution is adjusted and $\Psi$ is recomputed for the corrected parameters until it can no longer be decreased. This scheme ensures convergence in typically 10--15 iterations.

\subsection{Implementation in ZDI codes}
\label{sec:codes}

The methodology outlined above is implemented in a family of Fortran codes that share the same forward-modelling framework (Sects.~\ref{sec:synthesis} and \ref{sec:grid}), magnetic field parameterisation (Sect.~\ref{sec:harmonic}), regularisation functionals, and optimisation algorithm (Sect.~\ref{sec:inverse}), but differ in the way the local Stokes parameter spectra are obtained. All of them reconstruct the magnetic field topology simultaneously with the surface distributions of one or several chemical elements, using an arbitrary combination of Stokes parameters in individual lines or LSD profiles.

{\sc Invers10} \citep{piskunov:2002a,kochukhov:2002c} is the original implementation of full Stokes vector ZDI. It solves the polarised radiative transfer equation on the fly for every surface element and every rotational phase, using the opacities precomputed for a single model atmosphere, and treats the local abundances of the mapped elements through the scaling of the corresponding line opacities. This approximation makes it possible to reconstruct the abundance distributions of several chemical elements simultaneously with the magnetic field topology, provided that suitable spectral lines are available for each of them. Since these radiative transfer calculations have to be repeated at every iteration, they dominate the computational cost of the inversion and are parallelised with the help of the standard Message Passing Interface (MPI) libraries. This code has been the main tool for full Stokes vector magnetic mapping of Ap stars \citep[e.g.][]{kochukhov:2004d,kochukhov:2010,silvester:2014,silvester:2015,rusomarov:2015}. Here we employed it for all inversions using individual \ion{Fe}{ii} spectral lines (Sect.~\ref{sec:test_indiv}).

{\sc Invers13} \citep{kochukhov:2012,kochukhov:2013} is a further development of {\sc Invers10} in which the scalar surface parameter, either chemical abundance or temperature, is treated self-consistently with the local model atmosphere structure. Instead of relying on a single atmosphere, it solves the radiative transfer equation for the grid models bracketing the local value of that parameter and interpolates between the resulting Stokes profiles, thereby accounting for the feedback of the local composition or temperature on the atmospheric structure and on both the line and the continuum opacity. Since \citet{kochukhov:2012} and \citet{kochukhov:2017} demonstrated that this refinement has a minor impact on the magnetic and abundance maps of Ap stars, we adopted the simplified multi-element treatment of {\sc Invers10} described above and did not use {\sc Invers13} in the numerical experiments reported below. The code is, however, indispensable for elements with a strong influence on the atmospheric structure, such as He, and for the temperature mapping of cool active stars.

{\sc InversLSD} \citep{kochukhov:2014} implements the LSD-based inversion described in Sect.~\ref{sec:lsd}. Its reconstruction procedure follows that of {\sc Invers13}, except that the local Stokes profiles are not computed on the fly but are obtained by linear interpolation within the precomputed table of local LSD profiles and continuum intensities. Since this table is calculated for a grid of model atmospheres, the self-consistency of {\sc Invers13} is retained. In contrast to {\sc Invers10} and {\sc Invers13}, the computational cost is thus shifted entirely to the construction of this table, which is carried out once for a given set of stellar parameters, ahead of any inversion. The inversion itself is inexpensive and requires no parallelisation. {\sc InversLSD} has been widely applied to the LSD profiles of early-type stars, both in Stokes $IV$ \citep[e.g.][]{kochukhov:2016a,kochukhov:2022,kochukhov:2023a,kochukhov:2025} and in all four Stokes parameters \citep[e.g.][]{kochukhov:2017a,kochukhov:2019}. Here we used this code for all inversions based on LSD profiles (Sect.~\ref{sec:test_lsd}).

\section{Numerical experiments}
\label{sec:experiments}

\subsection{Magnetic structure from MHD fossil field models}
\label{sec:mhd}

\begin{figure*}[!th]
\centering
\fifps{0.57\hsize}{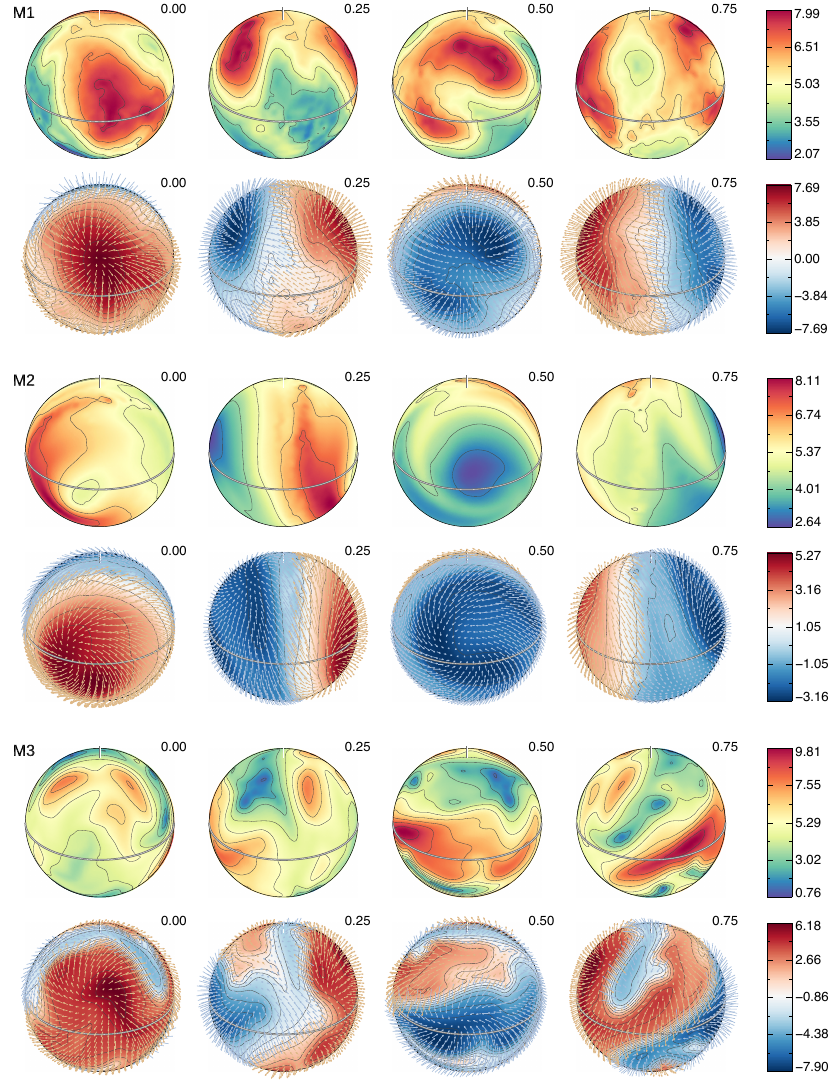}
\hspace*{0.5cm}
\fifps{0.33\hsize}{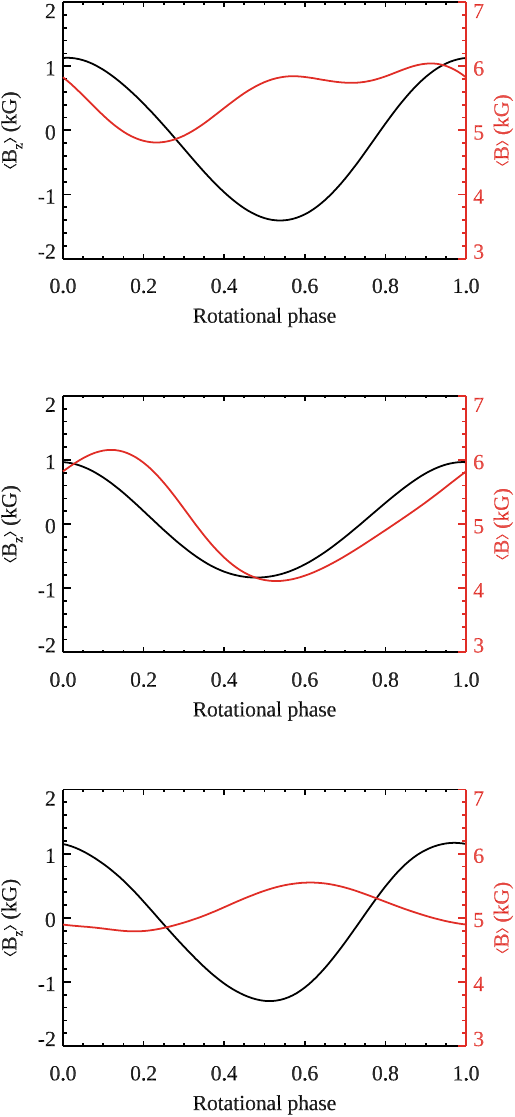}
\caption{\textit{Left:} Surface magnetic field topology of the three MHD models adopted for ZDI tests in this study. The star is shown at an inclination angle of $i=60\degr$ for the four rotation phases indicated next to each sphere. For each model (M1--M3) the upper row shows the field modulus and the lower one the radial field component and the field vector orientation. The thick line and the vertical bar mark the rotational equator and the pole, respectively. The contours are plotted with a 1~kG step and the colour bars give the field strength in kG.
\textit{Right:} Phase variation of the mean longitudinal magnetic field (black) and mean field modulus (red) for the same three models.
}
\label{fig:true}
\end{figure*}

The magnetic field topologies used in our numerical experiments were taken from the three-dimensional MHD simulations described by \citet{braithwaite:2004}, \citet{braithwaite:2006}, and \citet{braithwaite:2008}. These calculations follow the relaxation of an arbitrary, small-scale initial magnetic field in the stably stratified radiative interior of an intermediate-mass star. Within a few Alfv\'en crossing times such a field settles into a stable equilibrium configuration, with poloidal and toroidal components of comparable strength. Depending on the radial profile of the initial field strength, the relaxed configuration is either an approximately axisymmetric torus encircling the equator, which appears at the surface as a distorted dipole \citep{braithwaite:2006}, or a non-axisymmetric arrangement of flux tubes meandering below the surface, which yields a distinctly non-dipolar surface field \citep{braithwaite:2008}. These equilibria subsequently evolve only on the much longer Ohmic diffusion timescale.

The simulations neglect rotation and stellar winds, are not coupled to realistic stellar evolution calculations, and do not predict the absolute magnetic field strength. They nevertheless provide physically self-consistent, stable magnetic configurations of the kind that can plausibly exist in real Ap stars, which makes them far more realistic ZDI test cases than the low-order multipolar geometries prescribed analytically in earlier numerical experiments \citep{kochukhov:2002c}. We selected three of the resulting surface field maps, referred to as M1, M2, and M3 below, to represent the characteristic geometries encountered in ZDI studies of early-type stars. Model M1 is a typical distorted dipole, similar to the topologies reconstructed for many Ap stars \citep[e.g.][]{kochukhov:2010,silvester:2015,kochukhov:2019}. It corresponds to a deeply buried, slightly off-centre torus, accompanied by weaker current-carrying structures closer to the surface. Model M2 is likewise dipolar in terms of the dominant angular scale, but the underlying torus lies closer to the surface and over half of its surface magnetic energy is concentrated in the toroidal component. This type of field morphology was historically ignored by non-ZDI magnetic geometry studies of early-type stars \citep[e.g.][]{landstreet:2000,bagnulo:2002}, but a handful of Ap stars are known from ZDI to host substantial, if somewhat less dominant, toroidal fields \citep{oksala:2018,rusomarov:2018}. Model M3 has a markedly non-dipolar radial field distribution, comprising several regions of alternating polarity, and is thus reminiscent of the few early-type stars whose fields depart strongly from a dipole, such as the He-strong star HD\,37776 \citep{kochukhov:2011a} and the weak-field early-B star $\tau$~Sco \citep{donati:2006b,kochukhov:2016a,andersson:2025}. This surface pattern traces a system of twisted flux tubes meandering just below the surface: the radial field has opposite signs on the two sides of each tube, while the field strength peaks above the tubes and drops to low values between them.

Each map was interpolated onto the surface grid of 1876 elements employed for the disk integration and inversions (Sect.~\ref{sec:grid}) and scaled to a surface-averaged field modulus of 5~kG, which is typical of the strong-field Ap stars accessible to full Stokes vector ZDI. The maps were oriented with the axis of the underlying magnetic structure roughly perpendicular to the stellar rotation axis. Such large obliquities are common among Ap stars with rotation periods shorter than about 25~d \citep{landstreet:2000} and maximise the rotational modulation of the Stokes parameter profiles. 

\begin{table}[!t]
\centering
\caption{Characteristics of MHD magnetic field maps. 
\label{tbl:maps}}
\begin{tabular}{lccc}
\hline\hline
 & \multicolumn{3}{c}{Field model} \\
Parameter & M1 & M2 & M3 \\
\hline
$B_{\rm min}$ (kG) & 2.066 & 2.653 & 0.779 \\
$B_{\rm max}$ (kG) & 7.983 & 8.097 & 9.786 \\[3pt]
$E_{\rm pol}$ (\%) & 90.1 & 38.1 & 72.7 \\
$E_{\rm tor}$ (\%) &  9.9 & 61.9 & 27.3 \\[3pt]
$E_{\ell=1}$ (\%)  & 81.0 & 90.4 & 55.6 \\
$E_{\ell=2}$ (\%)  & 13.8 &  6.3 &  9.6 \\
$E_{\ell=3}$ (\%)  &  1.4 &  2.0 & 13.0 \\
$E_{\ell=4}$ (\%)  &  2.3 &  0.8 &  7.2 \\
$E_{\ell=5}$ (\%)  &  0.5 &  0.2 &  6.1 \\
$E_{\ell=6\mbox{--}10}$ (\%)  &  1.0 &  0.3 &  8.5 \\
\hline
\end{tabular}
\end{table}

The three field topologies are illustrated in the left panel of Fig.~\ref{fig:true} and characterised in Table~\ref{tbl:maps}. The latter lists the minimum and maximum local field modulus of each map, the fractional contributions of the poloidal and toroidal field components to the surface-integrated magnetic energy, and the fraction of that energy associated with the individual angular degrees from $\ell=1$ to 5 as well as with the $\ell=6$ to 10 modes taken together. These energies were obtained by expanding the maps into the harmonic series of Sect.~\ref{sec:harmonic} up to $\ell_{\rm max}=10$\footnote{A higher-order expansion yields $E_{\ell >10} = 1.8$\% for M3 and negligible contributions for the other two field maps.}. 

In the MHD simulations the region above the stellar surface is represented by a low-density, highly resistive medium, which serves as a computationally convenient substitute for a vacuum and allows the magnetic field to relax to a nearly current-free configuration \citep{braithwaite:2006,braithwaite:2008}. The radial and horizontal poloidal components of the input maps are consequently linked, with $\alpha_{\ell,m}=\beta_{\ell,m}$, matching the constrained parameterisation discussed in Sect.~\ref{sec:harmonic}. This condition applies to the poloidal field alone: the toroidal component, which is maintained by currents inside the star and does not continue into the current-free exterior, is left unconstrained and remains substantial in the input maps.

For each of the three maps, the right panel of Fig.~\ref{fig:true} shows the phase curves of the mean longitudinal magnetic field \bz\ and the mean field modulus \bm, computed for the inclination angle $i=60\degr$ adopted in all our numerical experiments. Despite the complexity of the underlying surface distributions, all three models yield smooth variations of both quantities, with the longitudinal field following a nearly sinusoidal curve. At first glance these curves appear dipole-like and are entirely typical of those observed for magnetic Ap stars \citep[e.g.][]{mathys:1991,mathys:2017}.

\subsection{Setup for numerical experiments}
\label{sec:setup}

All our numerical experiments shared the same stellar and observational configuration, chosen to represent a moderately rotating Ap star observed with a modern high-resolution spectropolarimeter. The local Stokes spectra were computed with an {\sc LLmodels} \citep{shulyak:2004} model atmosphere (\teff\ $=11\,000$~K, \lgg\ $=4.0$, scaled solar metallicity $+0.5$~dex) and a uniform Fe abundance of $\log (N_{\rm Fe}/N_{\rm tot})=-4.0$. The star was viewed at an inclination angle of $i=60\degr$, with a zero azimuth angle of the rotational axis and a projected rotational velocity of \vs\ $=30$~\kms. Simulated observations were generated for 20 equidistant rotational phases and convolved with a Gaussian instrumental profile corresponding to a resolving power of $R=65\,000$, matching the ESPaDOnS and NARVAL spectropolarimeters that have provided most of the four Stokes parameter data of early-type stars. This choice of stellar parameters and observational data is close to optimal for ZDI. The consequences of a lower data quality, gaps in the phase coverage, and inaccurate stellar parameters were examined by \citet{kochukhov:2002c} for simpler magnetic geometries. Here we deliberately adopted favourable conditions and inverted with the exact input values of $i$, the azimuth angle, and \vs, so that the reconstruction errors reflect the complexity of the underlying magnetic field rather than deficiencies of the data or errors in the stellar parameters.

The first series of experiments followed the classical individual-line full Stokes vector inversion approach, using {\sc Invers10} and the three \ion{Fe}{ii} lines of multiplet 42 listed in Table~\ref{tbl:lines}. These are among the strongest metal lines in the spectra of A-type stars and are the first to develop detectable linear polarisation signatures \citep{wade:2000b,silvester:2012}. This has made them a common choice for quantitative Stokes profile modelling \citep{bagnulo:2001,khalack:2006} and for the early ZDI studies of Ap stars \citep{kochukhov:2004d,kochukhov:2010}. Their strength is also a disadvantage: being saturated, these lines respond weakly to abundance inhomogeneities and are sensitive to the vertical abundance stratification. Random noise corresponding to S/N\,$=500$, representative of high-quality observations of bright Ap stars \citep{silvester:2012}, was added to their simulated spectra. For the field models studied here this is a factor of six to nine below the peak amplitude of the Stokes $Q$ and $U$ signatures of the three lines.

\begin{table}[!t]
\centering
\caption{\ion{Fe}{ii} spectral lines used in the individual-line ZDI calculations.
\label{tbl:lines}}
{\small
\setlength{\tabcolsep}{5pt}
\begin{tabular}{cccccccc}
\hline\hline
$\lambda$ (\AA) & $E_{\rm lo}$ (eV) & $\log gf$ & $J_{\rm l}$ & $J_{\rm u}$ & $g_{\rm l}$ & $g_{\rm u}$ & $g_{\rm eff}$ \\
\hline
4923.9212 &  2.891 & $-1.504$ & 2.5 & 1.5 & 2.00 & 2.40 & 1.70 \\
5018.4356 &  2.891 & $-1.345$ & 2.5 & 2.5 & 2.00 & 1.87 & 1.93 \\
5169.0282 &  2.891 & $-1.250$ & 2.5 & 3.5 & 2.00 & 1.70 & 1.33 \\
\hline
\end{tabular}
}
\tablefoot{Columns give the central wavelength, the excitation energy of the lower level, the oscillator strength, the $J$ numbers and the Land\'e factors of the lower and upper atomic levels, and the effective Land\'e factor, $g_{\rm eff}$. The line data are from VALD, with the oscillator strengths originating from \citet{raassen:1998}.}
\end{table}

The second series of experiments repeated the same tests with LSD profiles and {\sc InversLSD}. The setup was identical, except that the spectra were synthesised over the entire 4000--7000~\AA\ region, which contains more than 4000 lines, and the noise added to them corresponded to S/N\,$=200$. The LSD profiles were extracted with a mask of 968 \ion{Fe}{i} and \ion{Fe}{ii} lines whose central depths, evaluated before any broadening, exceed 5\% of the continuum. The resulting mean profiles reached a S/N of about $10\,000$. The same mask was used to compute the table of local LSD Stokes profiles (Sect.~\ref{sec:lsd}), with steps of 500~G in field strength and 0.2~dex in Fe abundance, and 15 values each of the limb angle and of the angle between the field vector and the line of sight.

For each field model and each type of input data we carried out three inversions: a full Stokes $IQUV$ reconstruction with the most general harmonic expansion, i.e. independent $\alpha_{\ell,m}$ and $\beta_{\ell,m}$; the same inversion using only Stokes $IV$; and a Stokes $IV$ inversion with the poloidal field constrained to $\alpha_{\ell,m}=\beta_{\ell,m}$. Since this constraint is satisfied by the input MHD maps (Sect.~\ref{sec:mhd}), the last experiment shows to what extent correct a priori information about the field geometry compensates for the information missing from circular polarisation data. In all cases the harmonic expansion was truncated at $\ell_{\rm max}=10$. All inversions started from a homogeneous Fe distribution and a zero magnetic field, and reconstructed the field topology and the abundance map simultaneously. Since the simulated observations correspond to a uniform Fe abundance, any structure recovered in the abundance map is spurious and measures the artefacts introduced by the cross-talk between magnetic and abundance reconstructions.

\begin{figure}[!t]
\centering
\fifps{0.85\hsize}{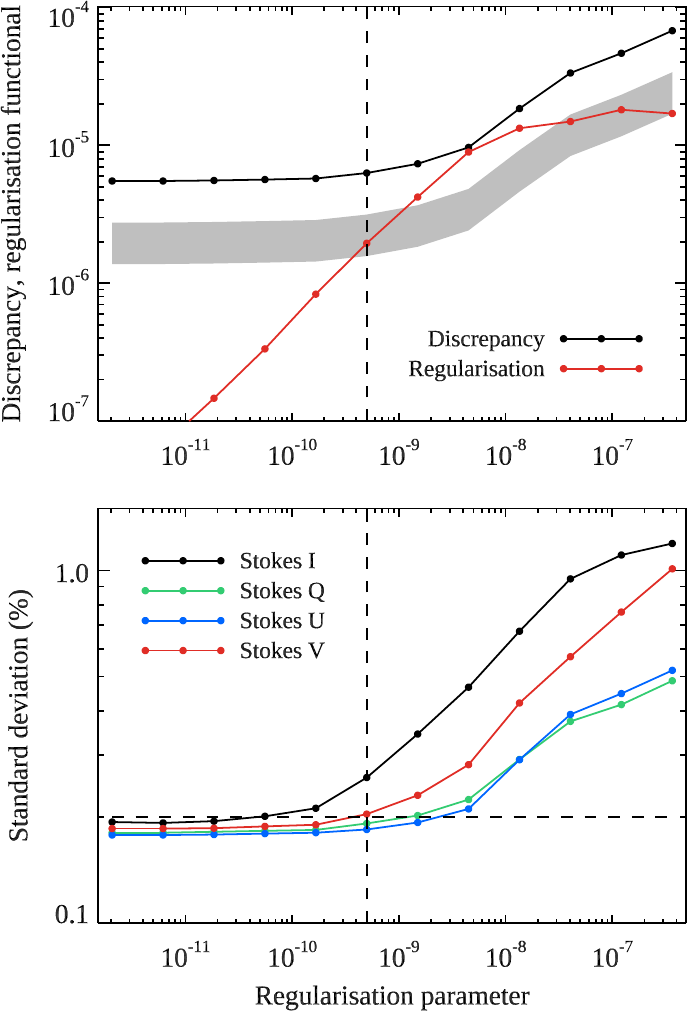}
\caption{Quality of the ZDI fit for the $IQUV$ M3 inversion as a function of the regularisation parameter $\Lambda_B$. \textit{Top}: Total discrepancy (black line) and regularisation $\Lambda_B R_B$ (red line) functionals, with the grey band showing the discrepancy functional divided by a factor of two to four. \textit{Bottom}: Standard deviation of the fit residuals in the individual Stokes parameters; the horizontal dashed line marks the noise level of the simulated data (S/N\,$=500$). In both panels, the vertical dashed line indicates our adopted $\Lambda_B$.}
\label{fig:regul}
\end{figure}

The magnetic regularisation parameter $\Lambda_B$ was calibrated with trial inversions spanning several orders of magnitude in its value, illustrated in Fig.~\ref{fig:regul} for the $IQUV$ inversion of model M3. Below a certain regularisation strength the discrepancy term of Eq.~(\ref{eq:psi}) stops improving while the regularisation term $\Lambda_B R_B$ keeps decreasing, indicating that the extra harmonic power admitted by the weaker constraint is not supported by the data. We therefore adopted the largest $\Lambda_B$ for which the regularisation term remained a factor of two to four below the discrepancy term, so that the solution was driven primarily by the observations without overfitting the noise. The lower panel of Fig.~\ref{fig:regul} confirms that this choice keeps the standard deviation of the fit residuals close to the injected noise level. This calibration was carried out separately for the individual-line and LSD experiments, with the same $\Lambda_B$ adopted for all three field models in each case; the results change very little upon recalibration for individual models and are only weakly sensitive to varying $\Lambda_B$ by a factor of two to three. The abundance regularisation parameter $\Lambda_{\varepsilon}$ was calibrated in the same way, except that its choice was guided by the fit to Stokes $I$, which primarily constrains the abundance map, whereas $\Lambda_B$ was set mainly by the fit to the polarisation profiles. Since the reconstructed maps are similarly insensitive to changes of $\Lambda_{\varepsilon}$ by a factor of a few, the two parameters were calibrated as independent one-dimensional problems rather than by a formal two-dimensional optimisation. Because our test cases contain no abundance inhomogeneities, the calibration of $\Lambda_{\varepsilon}$ in these experiments is not representative of abundance mapping of real stars, and we therefore do not discuss it in more detail.

\subsection{Characterisation of inversion errors}
\label{sec:errors}

We assess the quality of the reconstructions in two complementary ways: by comparing plots of the input and recovered maps, which shows where and in what form the inversion fails, and by quantifying the local differences between them, which allows different tests to be judged objectively and compared with each other. The differences were evaluated for the three magnetic field vector components, the field modulus, the field inclination $\zeta$ with respect to the surface normal, and the Fe abundance. The angle $\zeta$ is of particular interest because the orientation of the field lines relative to the surface governs the accumulation of chemical elements by atomic diffusion \citep{michaud:2015}.

Since different parts of the stellar surface contribute very unequally to the observed spectra, these differences have to be evaluated over a suitably restricted area. We quantified the contribution of a given latitude with a simple visibility function, obtained by averaging the projected surface area over all rotational phases. For the adopted inclination $i=60\degr$ this function vanishes below latitude $-60\degr$ and still reaches only 0.25 of its maximum at $-30\degr$. We therefore evaluated the error statistics for two limits: latitudes $\ge$\,$-60\degr$, comprising the entire surface visible at some rotational phase, and latitudes $\ge$\,$-30\degr$, retaining the reasonably well-observed part of the star.

\begin{figure*}[!t]
\sidecaption
\includegraphics[width=6.4cm]{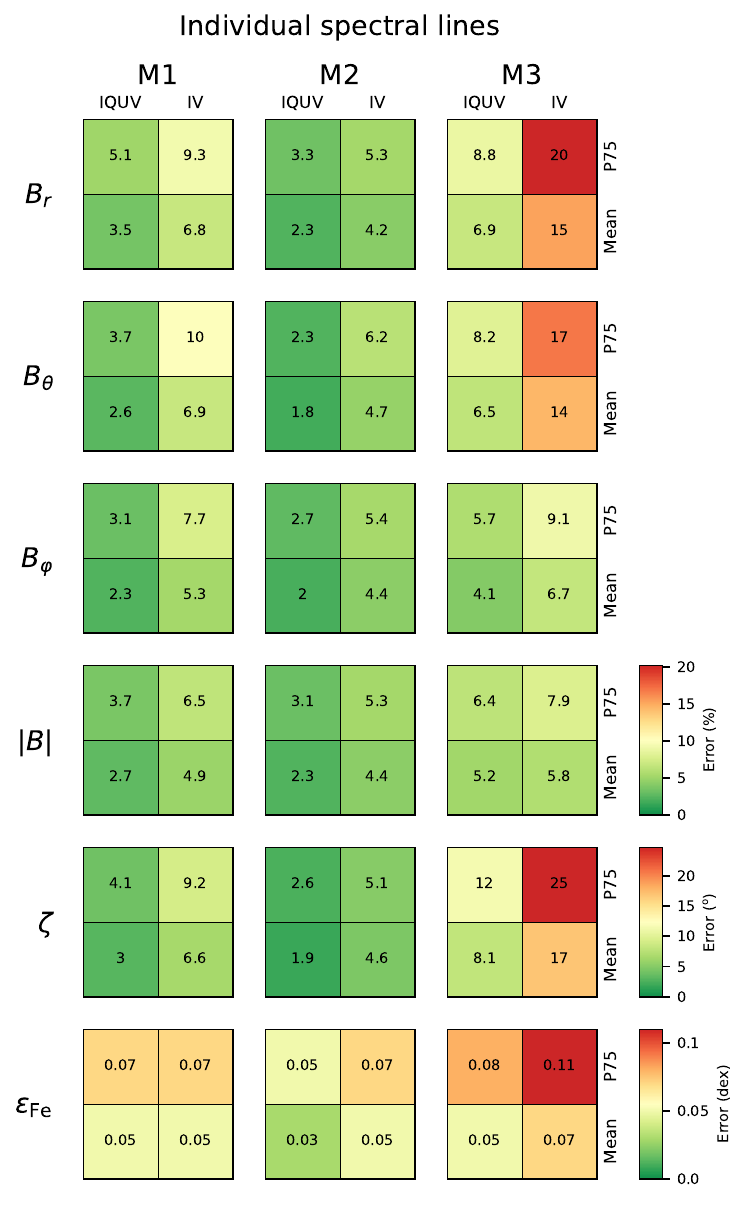}
\includegraphics[width=6.4cm]{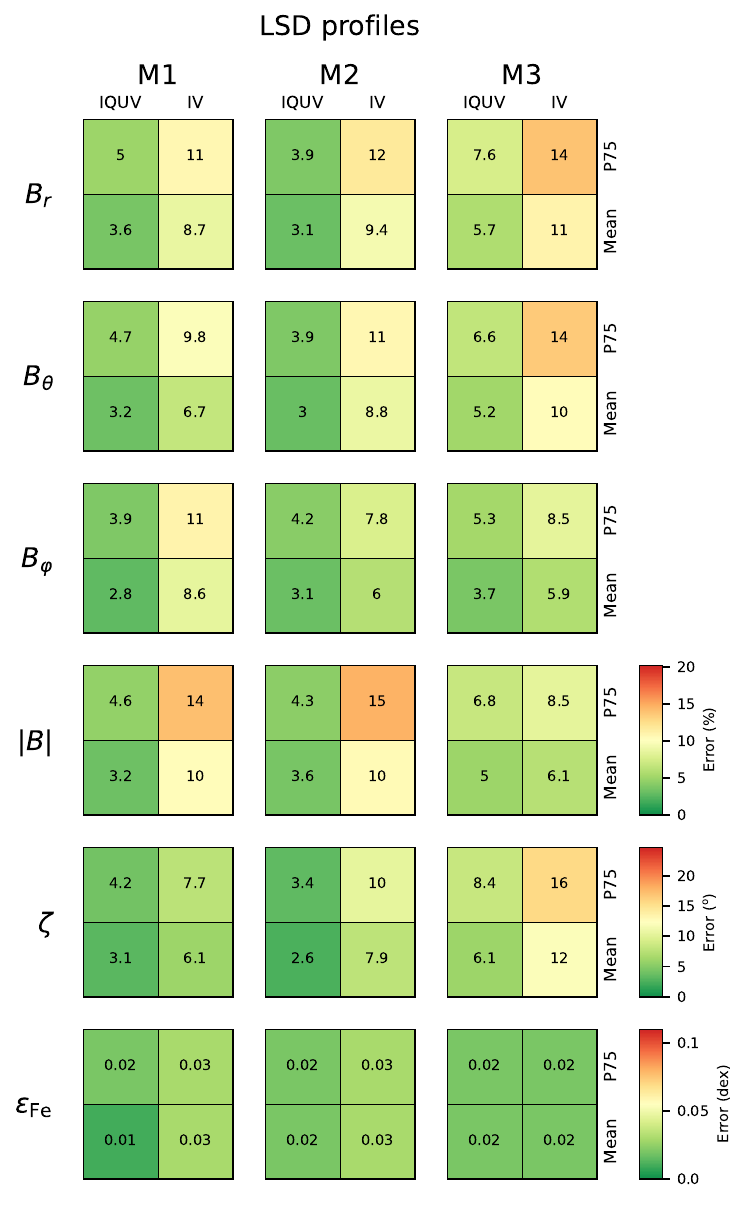}
\caption{Visual summary of the ZDI reconstruction errors obtained in our numerical tests with individual spectral lines (\textit{left}) and LSD profiles (\textit{right}). In each plot, the square symbols show the magnitude of the reconstruction error for the M1--M3 inversions (columns) and for the recovered surface distributions of the three magnetic vector components, field modulus, field inclination, and Fe abundance (rows). Each square is divided into four sectors: the left and right halves correspond to inversions based on the $IQUV$ and $IV$ profiles, respectively, while the upper and lower halves give the 75th-percentile and mean errors. The errors of the field components and field modulus are relative, expressed in per cent of $B_{\rm max}$, whereas those of the field inclination and abundance are given in degrees and dex, respectively, as indicated by the corresponding colour bars. In all cases we consider inversion results for the most general harmonic field parameterisation ($\alpha\ne\beta$) and for latitudes $\ge -30\degr$.}
\label{fig:summary}
\end{figure*}

For each quantity and each of these two areas we quote the mean absolute difference and the 75th percentile of the absolute differences, the latter characterising the upper envelope of the reconstruction errors. The errors of the field components and of the field modulus are expressed in per cent of the maximum field strength of the corresponding input map (Table~\ref{tbl:maps}), those of the field inclination in degrees, and those of the Fe abundance in dex. All numerical results are collected in Tables~\ref{tbl:tests} and \ref{tbl:tests_lsd} for the inversions using individual spectral lines and LSD profiles, respectively. Figure~\ref{fig:summary} summarises the mean and 75th-percentile errors of both data types for latitudes $\ge$\,$-30\degr$ and the general $\alpha_{\ell,m}\ne\beta_{\ell,m}$ parameterisation.

For the visual comparison, Figs.~\ref{fig:m1}, \ref{fig:m2}, and \ref{fig:m3} show, for each field model, the input maps together with the maps recovered from the Stokes $IQUV$ and $IV$ data and the corresponding difference maps, as well as the distribution of the magnetic energy over the angular degree $\ell$ and its poloidal and toroidal contributions. The same comparison for the LSD inversions is presented in Figs.~\ref{fig:m1_lsd}--\ref{fig:m3_lsd}, and the spurious Fe abundance distributions are shown in Figs.~\ref{fig:abn} and \ref{fig:abn_lsd}. All these figures correspond to the $\alpha_{\ell,m}\ne\beta_{\ell,m}$ inversions; the results obtained in the Stokes $IV$ inversions with the constrained parameterisation are reported in the tables only.

\subsubsection{ZDI with individual spectral lines}
\label{sec:test_indiv}

\begin{figure*}[!th]
\centering
\fihps{7.8cm}{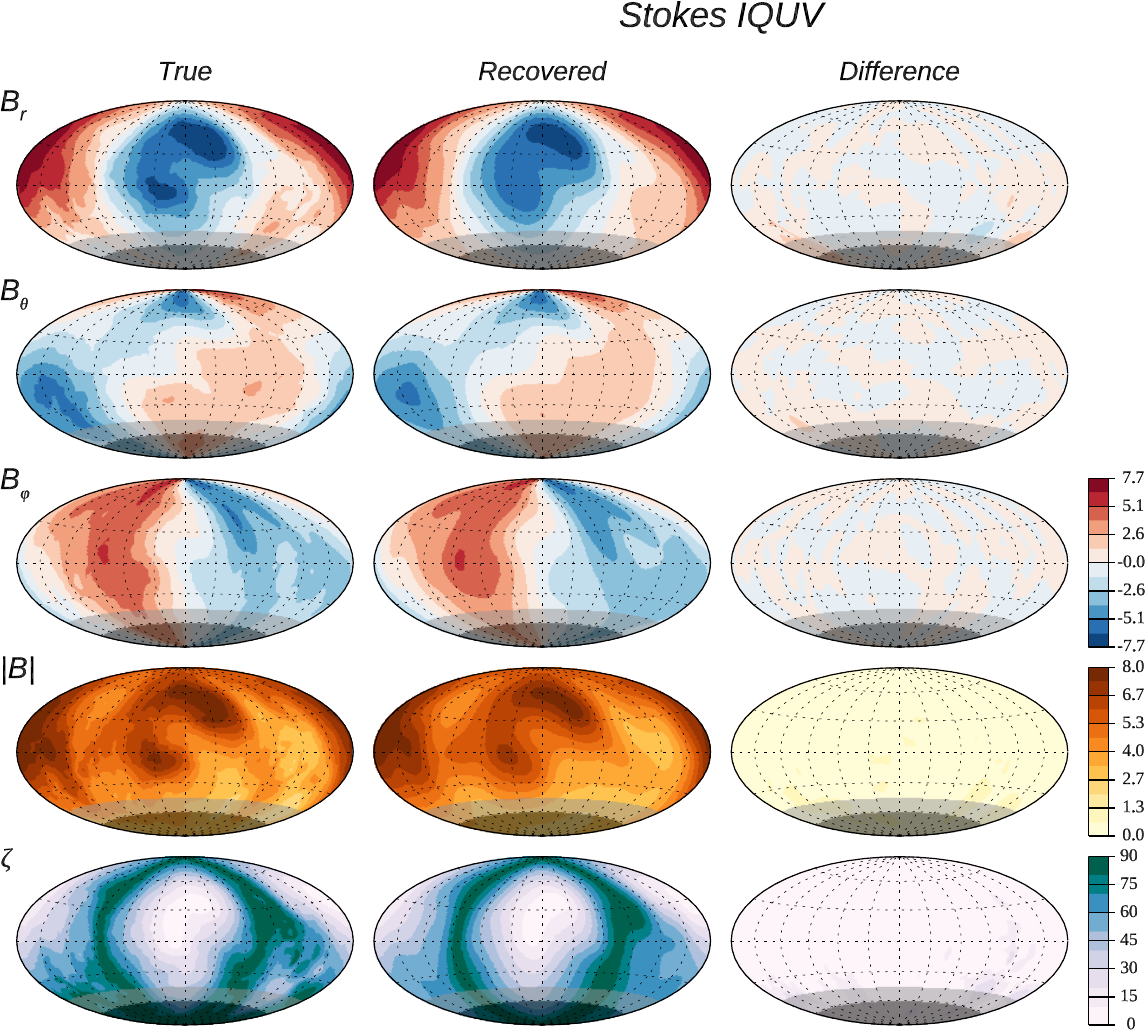}
\fihps{7.8cm}{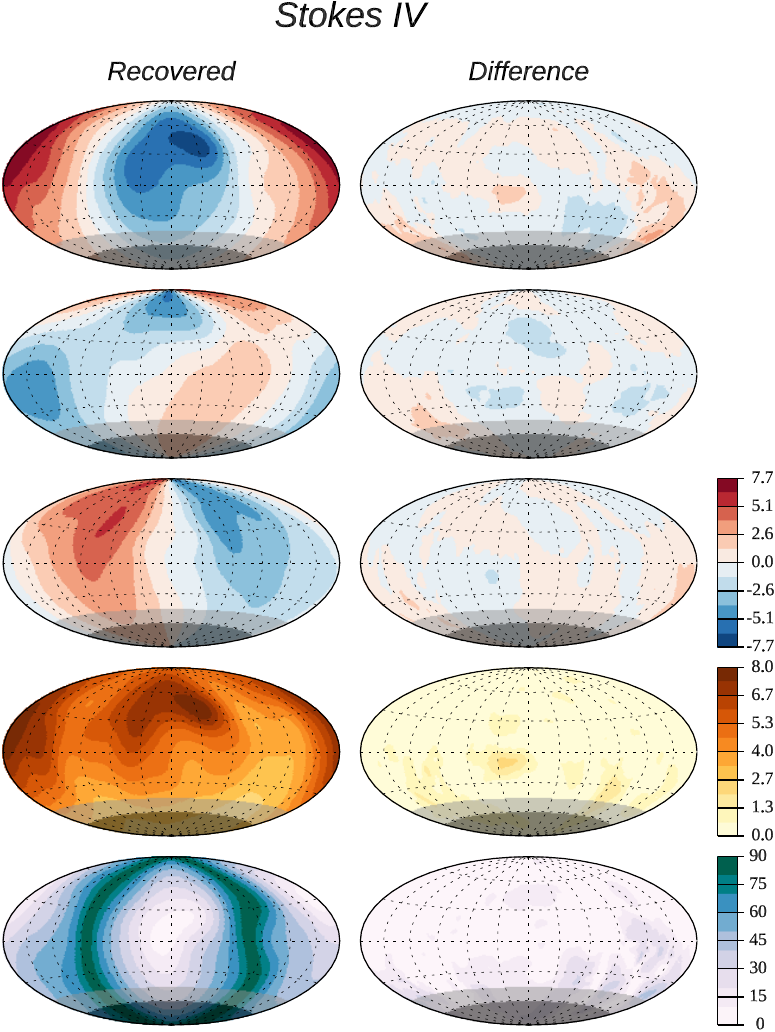}
\hspace*{0.2cm}
\fifps{3.2cm}{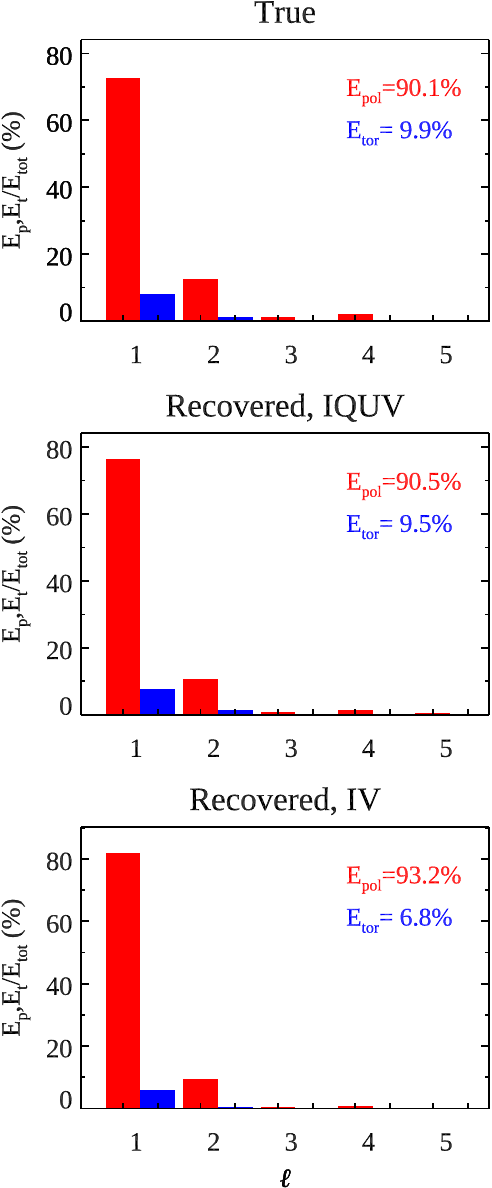}
\caption{Comparison of the true and reconstructed magnetic field distributions for the field geometry of M1. The columns show, from left to right, the true magnetic field maps, then the maps reconstructed in the $IQUV$ inversion and the corresponding difference maps (recovered minus true), and finally the same pair of maps for the $IV$ inversion. The rows show the radial, meridional, and azimuthal field components, the field modulus, and the field inclination with respect to the surface normal. The shaded bottom part of the Hammer--Aitoff maps indicates latitudes that are completely invisible (dark grey) or have a relative visibility of $<$\,0.1 (light grey). The three panels on the right compare the true and recovered distributions of the magnetic field energy as a function of the angular degree of the spherical harmonic modes, with the red and blue bars showing the poloidal and toroidal contributions, respectively.}
\label{fig:m1}
\end{figure*}

The full Stokes vector inversions reproduce all three input topologies closely (Figs.~\ref{fig:m1}, \ref{fig:m2}, and \ref{fig:m3}; Table~\ref{tbl:tests}). For latitudes $\ge$\,$-30\degr$ the mean errors of the individual field vector components are 1.8--6.9\% of $B_{\rm max}$ and the corresponding 75th percentiles 2.3--8.8\%, while the field modulus and the field inclination are recovered with mean errors of 2.3--5.2\% and 1.9--8.1\degr, respectively. The distorted dipole M1 and the toroidal model M2 are reconstructed equally well, with component errors of 1.8--3.5\%. The non-dipolar model M3 is about twice as difficult (4.1--6.9\%), with its residuals concentrated at the sharp boundaries between regions of opposite polarity, as expected from the $\ell_{\rm max}$ truncation and from the regularisation, which penalises high-order modes. Extending the statistics to latitudes $\ge$\,$-60\degr$ increases these errors by 20--50\%, and by up to a factor of two for M2. Although the general parameterisation leaves $\alpha_{\ell,m}$ and $\beta_{\ell,m}$ independent, these inversions return nearly identical values of the two coefficient families, thus recovering the current-free character of the poloidal field of the input maps from the data alone.

Restricting the inversion to Stokes $IV$ roughly doubles the errors for M1 and M2, to 4.2--6.9\% for the field components, and degrades the reconstruction of M3 much more severely: its radial and meridional field components are recovered with mean errors of 15.3 and 14.5\% (75th percentiles of 20.2 and 16.9\%) and its field inclination with a mean error of 16.8\degr. Figure~\ref{fig:m3} shows that the recovered maps are visibly smoother than the input ones, with the small-scale detail of the radial and meridional field largely lost. The field modulus is a notable exception: even for M3 it is reproduced with a mean error of 5.8\%, so that the Stokes $IV$ inversion recovers the distribution of the field strength while misplacing the field vector orientation.

The constrained $\alpha_{\ell,m}=\beta_{\ell,m}$ parameterisation, which the input maps satisfy, does not improve the Stokes $IV$ results systematically. For M1 it makes them worse, the mean errors of the radial field and of the field modulus increasing from 6.8 and 4.9\% to 10.4 and 10.5\%, and for M2 the changes are modest and of both signs. For M3, on the other hand, the constraint halves the errors of the radial and meridional components (7.7 and 7.1\%) and of the field inclination (9.1\degr), bringing this Stokes $IV$ reconstruction close to the full Stokes one. The a priori constraint is therefore useful only where the unconstrained Stokes $IV$ inversion fails.

The same ranking is seen in the harmonic energy distributions shown in the right-hand panels of Figs.~\ref{fig:m1}, \ref{fig:m2}, and \ref{fig:m3}. The Stokes $IQUV$ inversions reproduce the poloidal and toroidal energy fractions to within 0.4, 3.9, and 7.3 percentage points for M1, M2, and M3, and the poloidal and toroidal energies of the individual angular degrees to within 4 points for M1 and M2 and 8 points for M3. The Stokes $IV$ inversions distort this distribution considerably more (3.1, 9.8, and 17.6 points): for M3 this corresponds to an overestimated toroidal contribution at the expense of the poloidal power at $\ell\ge3$. Finally, the spurious Fe abundance inhomogeneities (Fig.~\ref{fig:abn}) do not exceed 0.05 dex on average and 0.08 dex at the 75th percentile for the Stokes $IQUV$ inversions, increasing to 0.07 and 0.11 dex for Stokes $IV$, with the largest deviations occurring in the poorly visible southern regions.

\subsubsection{ZDI with LSD profiles}
\label{sec:test_lsd}

The inversions of the LSD profiles behave in a qualitatively similar way (Figs.~\ref{fig:m1_lsd}--\ref{fig:m3_lsd}, Table~\ref{tbl:tests_lsd}). For the Stokes $IQUV$ data the mean errors of the field components are 2.8--5.7\%, those of the field modulus 3.2--5.0\%, and those of the field inclination 2.6--6.1\degr. These values are marginally worse than the individual-line results for the simple models M1 and M2 and better for M3, for which the radial field error decreases from 6.9 to 5.7\% and the inclination error from 8.1 to 6.1\degr. The recovery of the Fe abundance improves considerably: the spurious inhomogeneities do not exceed 0.03 dex in any of the LSD inversions (Fig.~\ref{fig:abn_lsd}).

For Stokes $IV$ the same pattern is more pronounced. The LSD inversions of M1 and M2 are less accurate than their individual-line counterparts, with component errors of 6.0--9.4\% and field modulus errors of up to 10.5\%, whereas M3 is recovered better (5.9--11.0\% for the components and 12.1\degr\ for the field inclination, compared with up to 15.3\% and 16.8\degr). The Stokes $IV$ inversions based on LSD profiles are thus somewhat less precise for simple field topologies, but more robust for complex ones.

The constrained parameterisation is more beneficial here than for the individual lines. It leaves the overall quality of the M1 reconstruction unchanged, with individual quantities moving in both directions, but improves M2 substantially, the mean errors of the radial field and of the field modulus dropping from 9.4 and 10.5\% to 4.9 and 3.6\%, and M3 as well, for which the radial field and inclination errors decrease to 6.8\% and 7.2\degr. Together with the results of Sect.~\ref{sec:test_indiv}, this indicates that linking the poloidal field components is worthwhile whenever the unconstrained Stokes $IV$ inversion performs poorly, and neutral or harmful otherwise.

The harmonic energy distribution is recovered with an accuracy similar to that of the individual-line inversions: the poloidal and toroidal fractions are reproduced to within 0.2--4.7 percentage points for Stokes $IQUV$ and 3.1--13.1 points for Stokes $IV$. Even when these totals are nearly correct, the energy can be redistributed between the angular degrees. For M1, for instance, the Stokes $IV$ inversion underestimates the poloidal energy of the $\ell=1$ modes by 11 points and overestimates that of the $\ell=2$ modes by 15, although the total poloidal fraction is too high by only 3 points.

\section{Conclusions and discussion}
\label{sec:conclusions}

We have used three surface magnetic field maps from three-dimensional MHD simulations of fossil fields to generate synthetic four Stokes parameter observations of an early-type star and to recover the underlying topologies with modern ZDI codes. Over the whole set of experiments the reconstructions are accurate: the field vector components and the field modulus are typically recovered to within about 10\% of the maximum field strength and the local field inclination to within about 10\degr, both for individual spectral lines and for LSD profiles. Errors of this magnitude, obtained for topologies far more complex than the low-order multipoles used in earlier tests, should be considered a success of the technique. The only clearly problematic case is the recovery of the most complex model M3 from Stokes $IV$ data alone, where the errors of the field components and of the inclination reach 15--20\% and 15--20\degr, respectively, in the individual-line inversions. Inversions of LSD profiles perform comparably to those based on individual \ion{Fe}{ii} lines and reproduce the (uniform) Fe abundance distribution appreciably better, which is important in practice, since LSD brings four Stokes parameter ZDI within reach for much fainter stars. LSD profiles also appear to be the more robust choice for difficult geometries: in the M3 Stokes $IV$ reconstruction they reduce the errors to 11\% and 12\degr. This advantage is absent for the simpler models M1 and M2 and is therefore unlikely to stem from the higher S/N of the LSD profiles. Instead, it likely reflects a difference in spatial resolution. For field strengths of 0--5~kG, the local profiles of the strong, Zeeman-sensitive \ion{Fe}{ii} lines are 9--19~\kms\ wide, compared with 6.5--13~\kms\ for the LSD profiles, which are dominated by weaker lines with smaller Zeeman splitting. Relative to \vs\ $=30$~\kms, the narrower LSD profiles resolve finer surface structures, which matters most for the highly structured map M3.

To our knowledge, this is the first time that self-consistent MHD models of fossil fields have been used to generate spectropolarimetric observables of early-type stars, and the first assessment of the ZDI performance for smaller-scale magnetic structures on these stars. Our tests also benchmark the ability of ZDI to recover the poloidal and toroidal components of a global fossil field simultaneously. These results are reassuring for the growing number of studies quantifying the toroidal energy content of early-type stars: the four Stokes parameter inversions reproduce the fractional poloidal and toroidal energies to within a few percentage points, and even the Stokes $IV$ inversions preserve the qualitative division between the two families.

A related result concerns the harmonic parameterisation itself. Although the general expansion allows the radial and horizontal poloidal components to vary independently, the Stokes $IQUV$ inversions converge to nearly identical $\alpha_{\ell,m}$ and $\beta_{\ell,m}$. Unconstrained full Stokes inversions therefore return a poloidal field consistent with a current-free exterior, and this property of the input maps is recovered from the data rather than imposed.

Enforcing $\alpha_{\ell,m}=\beta_{\ell,m}$ is a different matter. This constraint improves the Stokes $IV$ reconstructions precisely where the unconstrained inversion performs poorly, while degrading them for the simplest topology. Its usefulness rests, however, on an assumption satisfied by our input maps by construction, since the MHD simulations relax the field above the stellar surface to a nearly current-free state. As discussed in Sect.~\ref{sec:harmonic}, real stellar magnetospheres need not comply: an outflowing wind that opens the field beyond a finite radius unlinks the radial and horizontal poloidal components. This constraint should therefore be used to stabilise Stokes $IV$ inversions of complex fields, not as a default choice or as evidence that the underlying field is current-free.

It is instructive to compare the reconstruction errors summarised above with the empirical accuracy estimates obtained by reconstructing the field of the same star independently from different chemical elements. Such comparisons yield mean differences of 6--13\% of the peak field strength for the field components and modulus, and of 8--14\degr\ for the field inclination, both for four Stokes parameter inversions of weak-field stars \citep{kochukhov:2019} and for Stokes $IV$ reconstructions of much stronger fields \citep{kochukhov:2022,kochukhov:2025}. These estimates are of the same order as the errors established here, which is encouraging given that the two assessments are entirely independent. The agreement should not be pushed too far, however: differences between maps of different elements also absorb the numerous complications of real observations and real stars that our idealised experiments do not reproduce.

Finally, our experiments were carried out under deliberately favourable conditions: dense phase coverage, an intermediate inclination, moderate rotational broadening, and exact stellar parameters. The errors reported here therefore represent the best performance that current ZDI implementations can deliver for fossil fields of this complexity, and provide quantitative benchmarks for judging the reliability of published reconstructions.

\begin{acknowledgements}
O.K. acknowledges support from the Swedish Research Council (grant agreement no. 2023-03667) and the Swedish National Space Agency.
\end{acknowledgements}

\begin{appendix}

\onecolumn

\section{Additional figures and tables}

\begin{table*}[!th]
\centering
\caption{Reconstruction errors of ZDI inversions based on individual spectral lines. 
\label{tbl:tests}}
{\small
\begin{tabular}{ccccc|ccccc}
\hline\hline
Field   & Stokes         & Harmonic & \multicolumn{2}{c|}{Mean error/75th percentile}  & Field   & Stokes         & Harmonic & \multicolumn{2}{c}{Mean error/75th percentile} \\
model & parameters & expansion & lat\,$\ge$\,$-60\degr$  & lat\,$\ge$\,$-30\degr$     & model & parameters & expansion & lat\,$\ge$\,$-60\degr$  & lat\,$\ge$\,$-30\degr$    \\
\hline 
\multicolumn{5}{c|}{Radial field $B_r$ (\%)} &                             \multicolumn{5}{c}{Field modulus $|B|$ (\%)} \\
M1 & $IQUV$ & $\alpha\ne\beta$ & $4.6/6.2$ & $3.5/5.1$ &             M1 & $IQUV$ & $\alpha\ne\beta$ & $3.3/4.4$ & $2.7/3.7$ \\
   & $IV  $ & $\alpha\ne\beta$ & $8.2/11.4$ & $6.8/9.3$ &               & $IV  $ & $\alpha\ne\beta$ & $5.3/7.4$ & $4.9/6.5$ \\
   & $IV  $ & $\alpha=\beta$   & $12.7/18.0$ & $10.4/12.7$ &            & $IV  $ & $\alpha=\beta$   & $13.2/20.1$ & $10.5/13.6$ \\[1mm]
M2 & $IQUV$ & $\alpha\ne\beta$ & $2.9/4.1$ & $2.3/3.3$ &             M2 & $IQUV$ & $\alpha\ne\beta$ & $4.5/4.1$ & $2.3/3.1$ \\
   & $IV  $ & $\alpha\ne\beta$ & $5.0/6.4$ & $4.2/5.3$ &                & $IV  $ & $\alpha\ne\beta$ & $8.1/8.9$ & $4.4/5.3$ \\
   & $IV  $ & $\alpha=\beta$   & $6.3/8.0$ & $5.1/6.7$ &                & $IV  $ & $\alpha=\beta$   & $6.9/7.3$ & $4.7/5.5$ \\[1mm]
M3 & $IQUV$ & $\alpha\ne\beta$ & $9.1/12.4$ & $6.9/8.8$ &            M3 & $IQUV$ & $\alpha\ne\beta$ & $6.9/9.0$ & $5.2/6.4$ \\
   & $IV  $ & $\alpha\ne\beta$ & $18.3/27.1$ & $15.3/20.2$ &            & $IV  $ & $\alpha\ne\beta$ & $6.8/9.4$ & $5.8/7.9$ \\
   & $IV  $ & $\alpha=\beta$   & $9.7/12.5$ & $7.7/10.1$ &              & $IV  $ & $\alpha=\beta$   & $6.8/9.0$ & $5.0/6.6$ \\[1mm]
\hline
\multicolumn{5}{c|}{Meridional field $B_\theta$ (\%)} &                         \multicolumn{5}{c}{Field inclination $\zeta$ (\degr)} \\
M1 & $IQUV$ & $\alpha\ne\beta$ & $3.1/4.1$ & $2.6/3.7$ &             M1 & $IQUV$ & $\alpha\ne\beta$ & $4.4/5.7$ & $3.0/4.1$ \\
   & $IV  $ & $\alpha\ne\beta$ & $7.3/10.7$ & $6.9/10.0$ &              & $IV  $ & $\alpha\ne\beta$ & $8.9/12.4$ & $6.6/9.2$ \\
   & $IV  $ & $\alpha=\beta$   & $11.9/13.5$ & $7.4/9.4$ &              & $IV  $ & $\alpha=\beta$   & $11.6/14.7$ & $8.2/9.6$ \\[1mm]
M2 & $IQUV$ & $\alpha\ne\beta$ & $3.3/3.8$ & $1.8/2.3$ &             M2 & $IQUV$ & $\alpha\ne\beta$ & $2.8/3.6$ & $1.9/2.6$ \\
   & $IV  $ & $\alpha\ne\beta$ & $7.2/9.1$ & $4.7/6.2$ &                & $IV  $ & $\alpha\ne\beta$ & $7.0/8.7$ & $4.6/5.1$ \\
   & $IV  $ & $\alpha=\beta$   & $5.5/7.9$ & $3.6/4.7$ &                & $IV  $ & $\alpha=\beta$   & $5.0/7.5$ & $4.1/5.8$ \\[1mm]
M3 & $IQUV$ & $\alpha\ne\beta$ & $7.9/10.5$ & $6.5/8.2$ &            M3 & $IQUV$ & $\alpha\ne\beta$ & $10.3/13.7$ & $8.1/11.5$ \\
   & $IV  $ & $\alpha\ne\beta$ & $17.2/24.5$ & $14.5/16.9$ &            & $IV  $ & $\alpha\ne\beta$ & $19.6/29.9$ & $16.8/24.7$ \\
   & $IV  $ & $\alpha=\beta$   & $8.4/11.7$ & $7.1/9.8$ &               & $IV  $ & $\alpha=\beta$   & $11.1/14.9$ & $9.1/12.6$ \\[1mm]
\hline
\multicolumn{5}{c|}{Azimuthal field $B_\varphi$ (\%)} &                          \multicolumn{5}{c}{Element abundance $\varepsilon_{\rm Fe}$ ($\log(N_{\rm Fe}/N_{\rm tot})$)} \\
M1 & $IQUV$ & $\alpha\ne\beta$ & $2.8/3.7$ & $2.3/3.1$ &             M1 & $IQUV$ & $\alpha\ne\beta$ & $0.07/0.12$ & $0.05/0.07$ \\
   & $IV  $ & $\alpha\ne\beta$ & $5.7/8.4$ & $5.3/7.7$ &                & $IV  $ & $\alpha\ne\beta$ & $0.06/0.08$ & $0.05/0.07$ \\
   & $IV  $ & $\alpha=\beta$   & $11.7/14.5$ & $7.9/9.4$ &              & $IV  $ & $\alpha=\beta$   & $0.10/0.14$ & $0.08/0.11$ \\[1mm]
M2 & $IQUV$ & $\alpha\ne\beta$ & $3.7/3.6$ & $2.0/2.7$ &             M2 & $IQUV$ & $\alpha\ne\beta$ & $0.04/0.07$ & $0.03/0.05$ \\
   & $IV  $ & $\alpha\ne\beta$ & $7.3/7.3$ & $4.4/5.4$ &                & $IV  $ & $\alpha\ne\beta$ & $0.06/0.09$ & $0.05/0.07$ \\
   & $IV  $ & $\alpha=\beta$   & $8.0/11.3$ & $6.0/8.4$ &               & $IV  $ & $\alpha=\beta$   & $0.06/0.09$ & $0.06/0.09$ \\[1mm]
M3 & $IQUV$ & $\alpha\ne\beta$ & $5.4/6.8$ & $4.1/5.7$ &             M3 & $IQUV$ & $\alpha\ne\beta$ & $0.06/0.08$ & $0.05/0.08$ \\
   & $IV  $ & $\alpha\ne\beta$ & $8.0/10.3$ & $6.7/9.1$ &               & $IV  $ & $\alpha\ne\beta$ & $0.08/0.12$ & $0.07/0.11$ \\
   & $IV  $ & $\alpha=\beta$   & $6.8/9.3$ & $4.9/6.7$ &                & $IV  $ & $\alpha=\beta$   & $0.09/0.13$ & $0.08/0.12$ \\
\hline
\end{tabular}
}
\tablefoot{Columns 1 and 6 identify the field geometry model adopted for the ZDI tests. Columns 2 and 7 indicate the Stokes parameters used in the magnetic inversion, and columns 3 and 8 the adopted harmonic parameterisation. Columns 4 and 9 give the mean and 75th percentile reconstruction errors for latitudes $\ge$\,$-60\degr$, and columns 5 and 10 those for latitudes $\ge$\,$-30\degr$. The different parts of the table, separated by horizontal and vertical lines, quantify reconstruction errors for the three magnetic field vector components and the local field modulus (both in per cent of the maximum field strength), the field inclination with respect to the surface normal (in degrees), and Fe abundance (in $\log(N_{\rm Fe}/N_{\rm tot})$ units).}
\end{table*}

\begin{table*}[!th]
\centering
\caption{Reconstruction errors of ZDI inversions based on LSD profiles. 
\label{tbl:tests_lsd}}
{\small
\begin{tabular}{ccccc|ccccc}
\hline\hline
Field   & Stokes         & Harmonic & \multicolumn{2}{c|}{Mean error/75th percentile}  & Field   & Stokes         & Harmonic & \multicolumn{2}{c}{Mean error/75th percentile} \\
model & parameters & expansion & lat\,$\ge$\,$-60\degr$  & lat\,$\ge$\,$-30\degr$     & model & parameters & expansion & lat\,$\ge$\,$-60\degr$  & lat\,$\ge$\,$-30\degr$    \\
\hline 
\multicolumn{5}{c|}{Radial field $B_r$ (\%)} &                             \multicolumn{5}{c}{Field modulus $|B|$ (\%)} \\                     
M1 & $IQUV$ & $\alpha\ne\beta$ & $4.3/5.8$ & $3.6/5.0$ &             M1 & $IQUV$ & $\alpha\ne\beta$ & $4.0/5.6$ & $3.2/4.6$ \\      
   & $IV  $ & $\alpha\ne\beta$ & $10.2/13.6$ & $8.7/10.7$ &             & $IV  $ & $\alpha\ne\beta$ & $13.6/22.3$ & $10.4/14.1$ \\  
   & $IV  $ & $\alpha=\beta$   & $8.4/12.1$ & $7.8/11.7$ &              & $IV  $ & $\alpha=\beta$   & $11.9/16.8$ & $10.1/12.9$ \\[1mm]
M2 & $IQUV$ & $\alpha\ne\beta$ & $4.2/4.9$ & $3.1/3.9$ &             M2 & $IQUV$ & $\alpha\ne\beta$ & $7.3/8.6$ & $3.6/4.3$ \\      
   & $IV  $ & $\alpha\ne\beta$ & $10.5/14.1$ & $9.4/11.7$ &             & $IV  $ & $\alpha\ne\beta$ & $16.4/25.8$ & $10.5/14.6$ \\  
   & $IV  $ & $\alpha=\beta$   & $6.8/10.1$ & $4.9/6.1$ &               & $IV  $ & $\alpha=\beta$   & $5.3/5.9$ & $3.6/4.6$ \\[1mm]      
M3 & $IQUV$ & $\alpha\ne\beta$ & $8.0/9.7$ & $5.7/7.6$ &             M3 & $IQUV$ & $\alpha\ne\beta$ & $7.0/9.1$ & $5.0/6.8$ \\      
   & $IV  $ & $\alpha\ne\beta$ & $12.6/17.4$ & $11.0/13.9$ &            & $IV  $ & $\alpha\ne\beta$ & $8.9/11.5$ & $6.1/8.5$ \\     
   & $IV  $ & $\alpha=\beta$   & $9.3/11.5$ & $6.8/9.1$ &               & $IV  $ & $\alpha=\beta$   & $7.5/10.0$ & $5.7/7.8$ \\[1mm]     
\hline
\multicolumn{5}{c|}{Meridional field $B_\theta$ (\%)} &                         \multicolumn{5}{c}{Field inclination $\zeta$ (\degr)} \\                  
M1 & $IQUV$ & $\alpha\ne\beta$ & $3.5/4.9$ & $3.2/4.7$ &             M1 & $IQUV$ & $\alpha\ne\beta$ & $4.1/5.5$ & $3.1/4.2$ \\         
   & $IV  $ & $\alpha\ne\beta$ & $8.1/11.5$ & $6.7/9.8$ &               & $IV  $ & $\alpha\ne\beta$ & $7.4/9.8$ & $6.1/7.7$ \\         
   & $IV  $ & $\alpha=\beta$   & $12.9/19.0$ & $10.8/14.1$ &            & $IV  $ & $\alpha=\beta$   & $11.5/16.4$ & $8.2/12.1$ \\[1mm]      
M2 & $IQUV$ & $\alpha\ne\beta$ & $5.0/5.9$ & $3.0/3.9$ &             M2 & $IQUV$ & $\alpha\ne\beta$ & $3.8/4.9$ & $2.6/3.4$ \\         
   & $IV  $ & $\alpha\ne\beta$ & $12.4/19.7$ & $8.8/10.7$ &             & $IV  $ & $\alpha\ne\beta$ & $9.0/12.6$ & $7.9/10.5$ \\       
   & $IV  $ & $\alpha=\beta$   & $6.8/8.6$ & $4.4/6.0$ &                & $IV  $ & $\alpha=\beta$   & $6.4/10.2$ & $4.9/5.9$ \\[1mm]        
M3 & $IQUV$ & $\alpha\ne\beta$ & $6.8/9.6$ & $5.2/6.6$ &             M3 & $IQUV$ & $\alpha\ne\beta$ & $9.3/11.1$ & $6.1/8.4$ \\        
   & $IV  $ & $\alpha\ne\beta$ & $11.8/16.3$ & $10.3/13.5$ &            & $IV  $ & $\alpha\ne\beta$ & $14.3/20.0$ & $12.1/15.6$ \\     
   & $IV  $ & $\alpha=\beta$   & $8.3/10.4$ & $6.2/8.5$ &               & $IV  $ & $\alpha=\beta$   & $10.1/12.2$ & $7.2/9.6$ \\[1mm]       
\hline
\multicolumn{5}{c|}{Azimuthal field $B_\varphi$ (\%)} &                          \multicolumn{5}{c}{Element abundance $\varepsilon_{\rm Fe}$ ($\log(N_{\rm Fe}/N_{\rm tot})$)} \\
M1 & $IQUV$ & $\alpha\ne\beta$ & $3.5/4.8$ & $2.8/3.9$ &             M1 & $IQUV$ & $\alpha\ne\beta$ & $0.01/0.02$ & $0.01/0.02$ \\          
   & $IV  $ & $\alpha\ne\beta$ & $10.3/14.4$ & $8.6/11.0$ &             & $IV  $ & $\alpha\ne\beta$ & $0.03/0.03$ & $0.03/0.03$ \\         
   & $IV  $ & $\alpha=\beta$   & $7.8/10.7$ & $6.5/9.3$ &               & $IV  $ & $\alpha=\beta$   & $0.03/0.03$ & $0.03/0.03$ \\[1mm]         
M2 & $IQUV$ & $\alpha\ne\beta$ & $5.5/5.9$ & $3.1/4.2$ &             M2 & $IQUV$ & $\alpha\ne\beta$ & $0.02/0.02$ & $0.02/0.02$ \\         
   & $IV  $ & $\alpha\ne\beta$ & $9.6/11.2$ & $6.0/7.8$ &               & $IV  $ & $\alpha\ne\beta$ & $0.03/0.03$ & $0.03/0.03$ \\         
   & $IV  $ & $\alpha=\beta$   & $7.2/10.1$ & $5.2/7.1$ &               & $IV  $ & $\alpha=\beta$   & $0.01/0.01$ & $0.01/0.01$ \\[1mm]         
M3 & $IQUV$ & $\alpha\ne\beta$ & $4.9/6.5$ & $3.7/5.3$ &             M3 & $IQUV$ & $\alpha\ne\beta$ & $0.02/0.02$ & $0.02/0.02$ \\         
   & $IV  $ & $\alpha\ne\beta$ & $7.0/8.8$ & $5.9/8.5$ &                & $IV  $ & $\alpha\ne\beta$ & $0.02/0.02$ & $0.02/0.02$ \\         
   & $IV  $ & $\alpha=\beta$   & $8.3/10.9$ & $5.8/8.0$ &               & $IV  $ & $\alpha=\beta$   & $0.01/0.01$ & $0.01/0.01$ \\         
\hline
\end{tabular}
}
\tablefoot{The meaning of the columns is the same as in Table~\ref{tbl:tests}.}
\end{table*}

\begin{figure*}[!th]
\centering
\fihps{7.8cm}{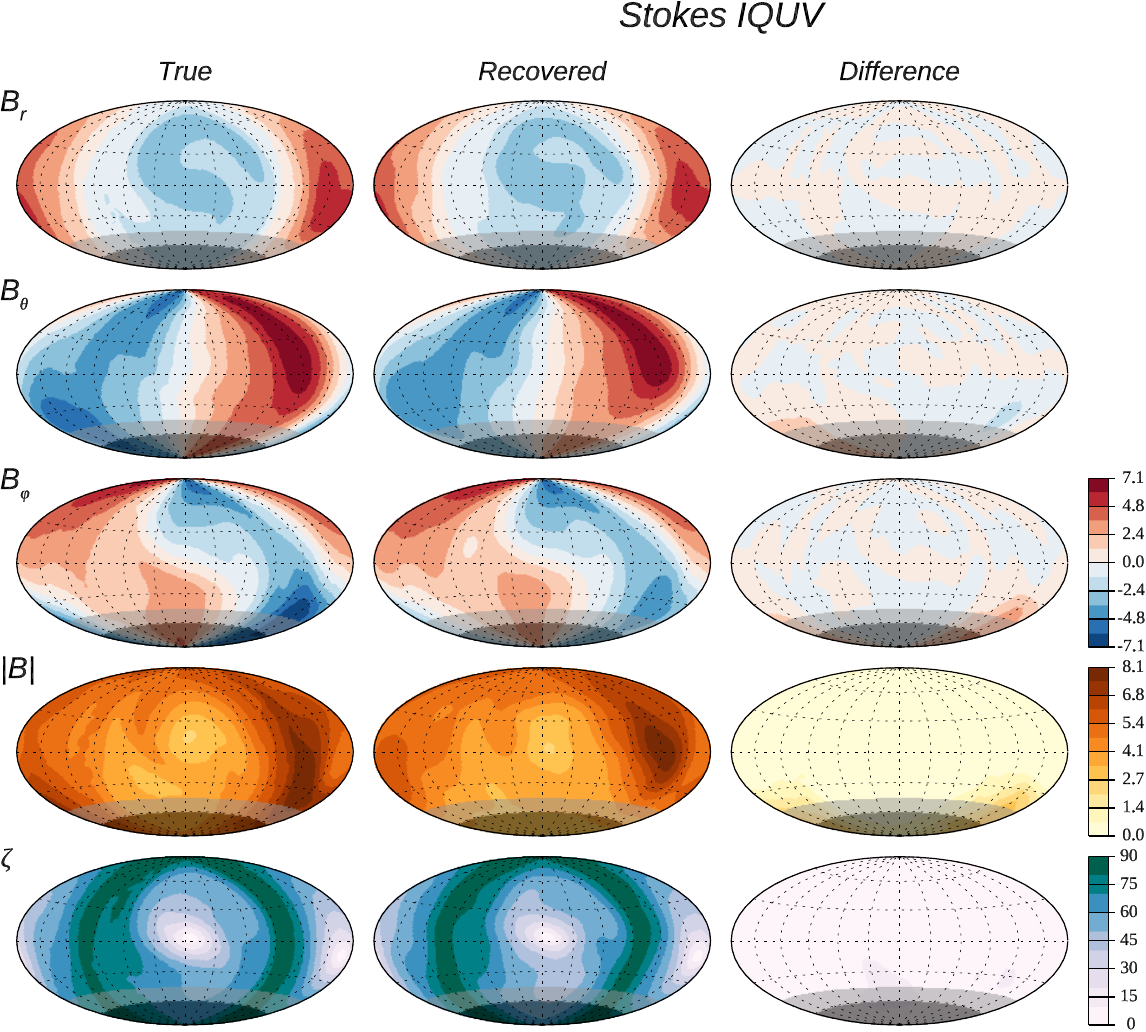}
\fihps{7.8cm}{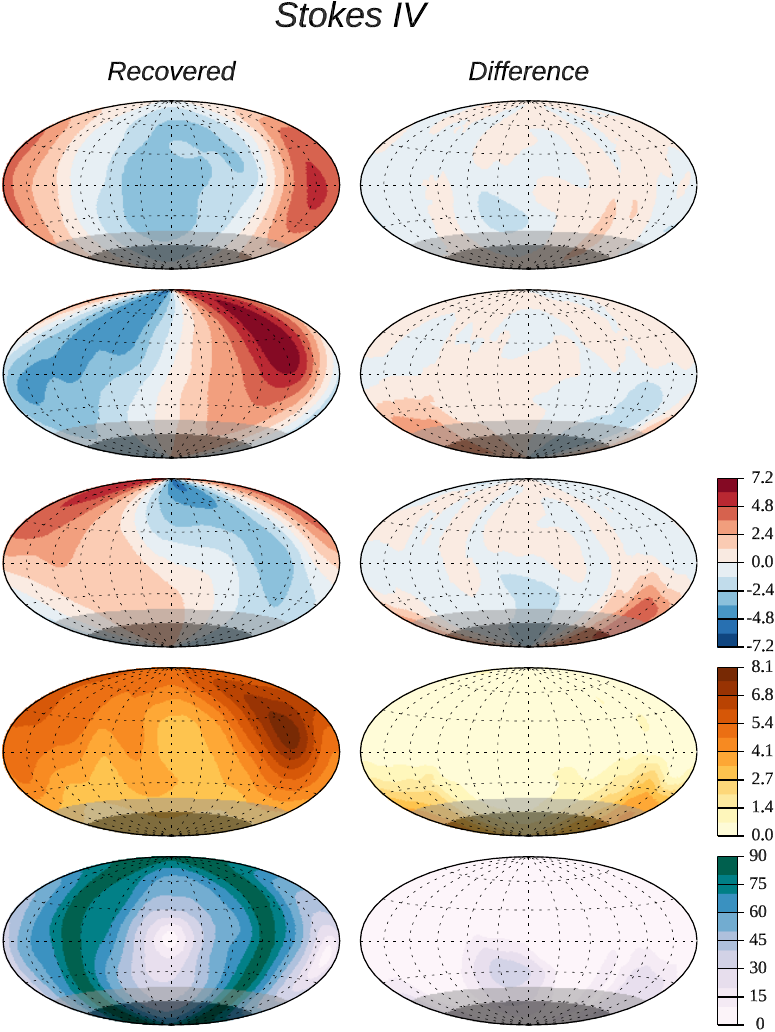}
\hspace*{0.2cm}
\fifps{3.2cm}{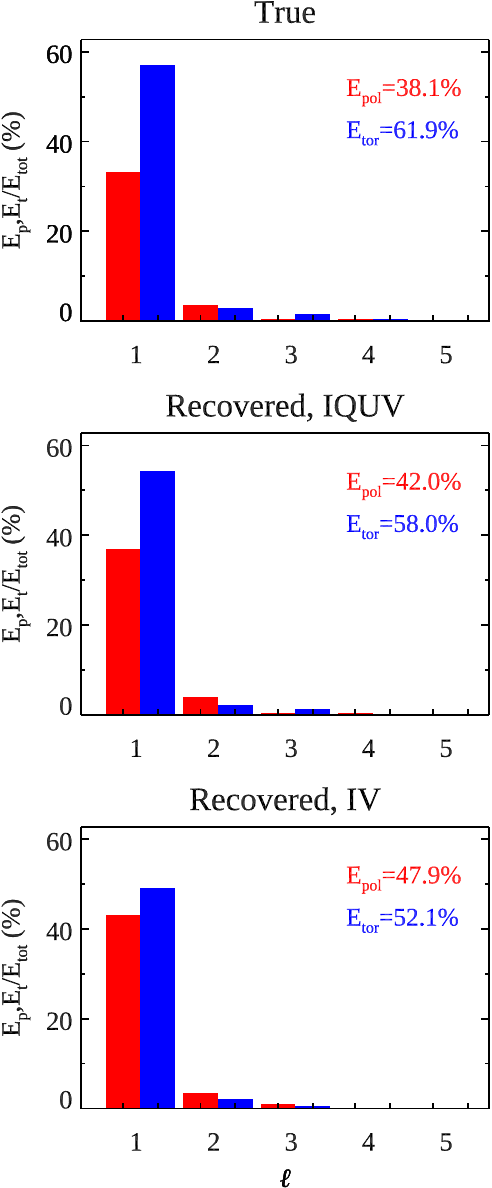}
\caption{Same as Fig.~\ref{fig:m1} but for ZDI reconstructions of model M2 using individual lines.}
\label{fig:m2}
\end{figure*}

\begin{figure*}[!th]
\centering
\fihps{7.8cm}{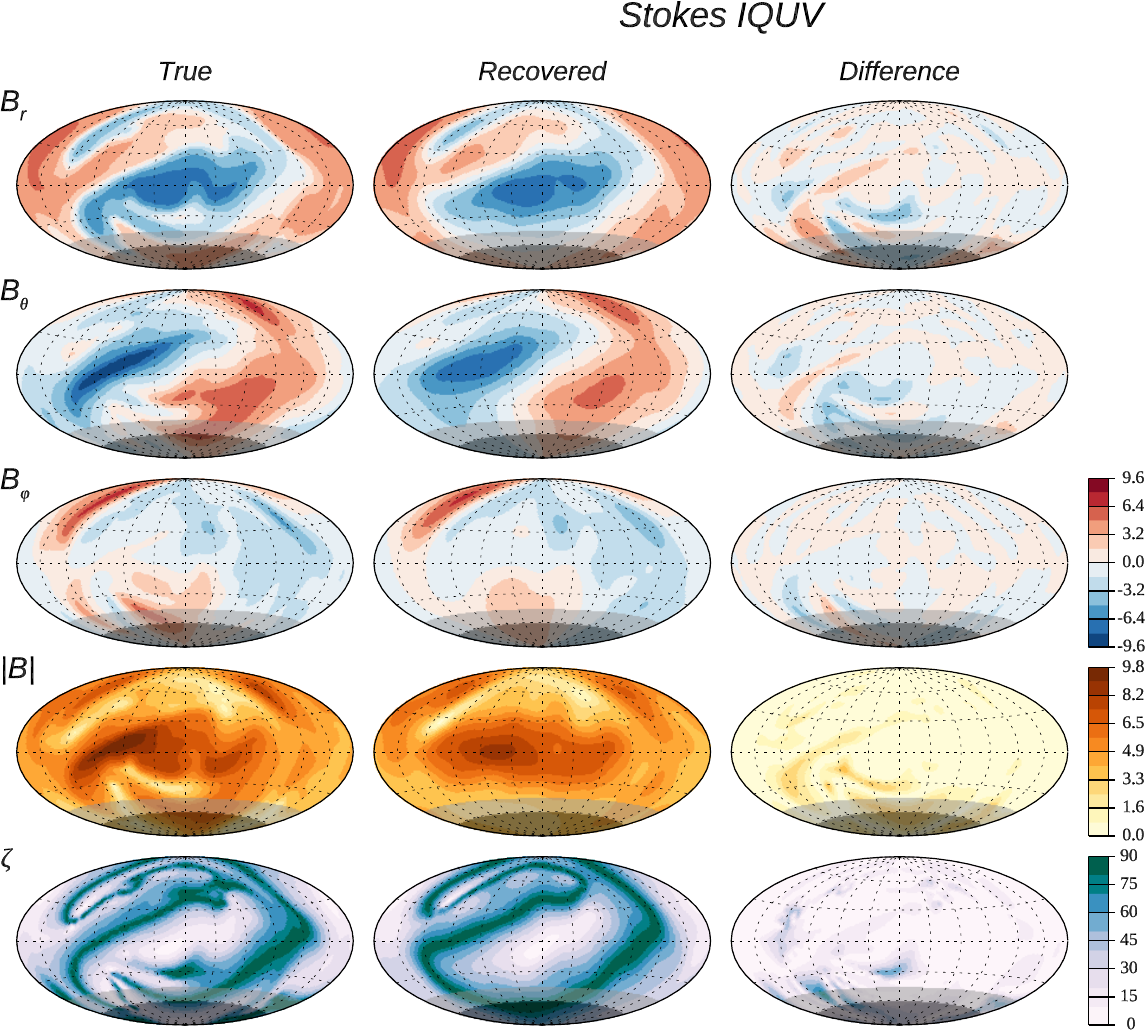}
\fihps{7.8cm}{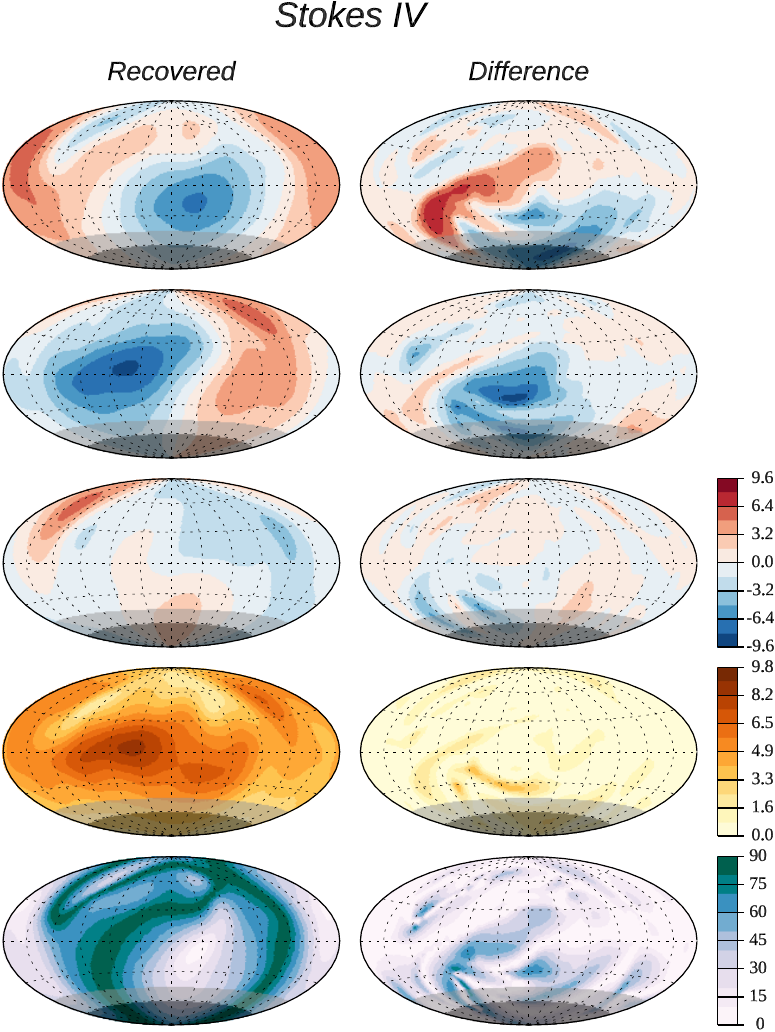}
\hspace*{0.2cm}
\fifps{3.2cm}{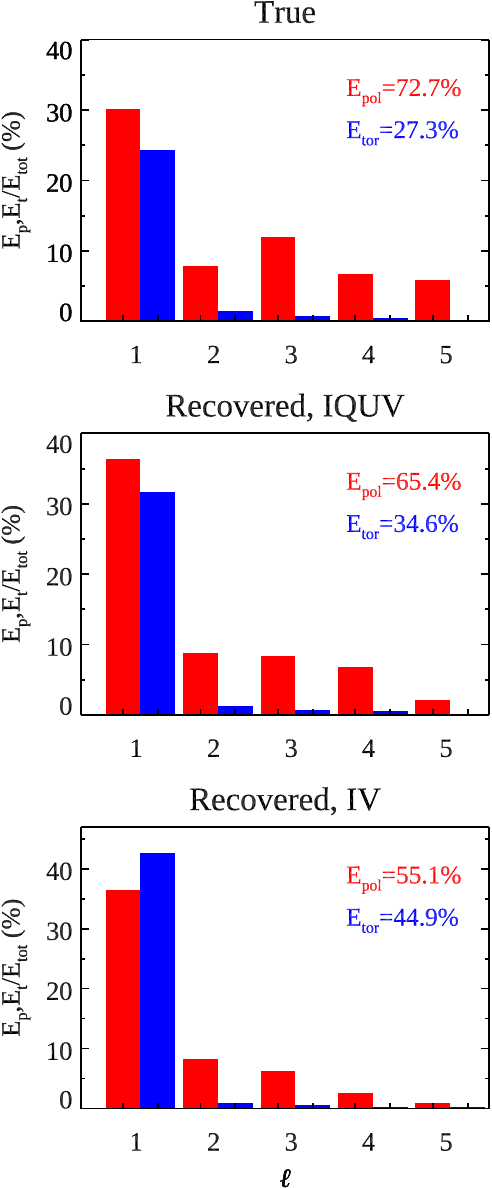}
\caption{Same as Fig.~\ref{fig:m1} but for ZDI reconstructions of model M3 using individual lines.}
\label{fig:m3}
\end{figure*}

\begin{figure*}[!th]
\centering
\fihps{7.8cm}{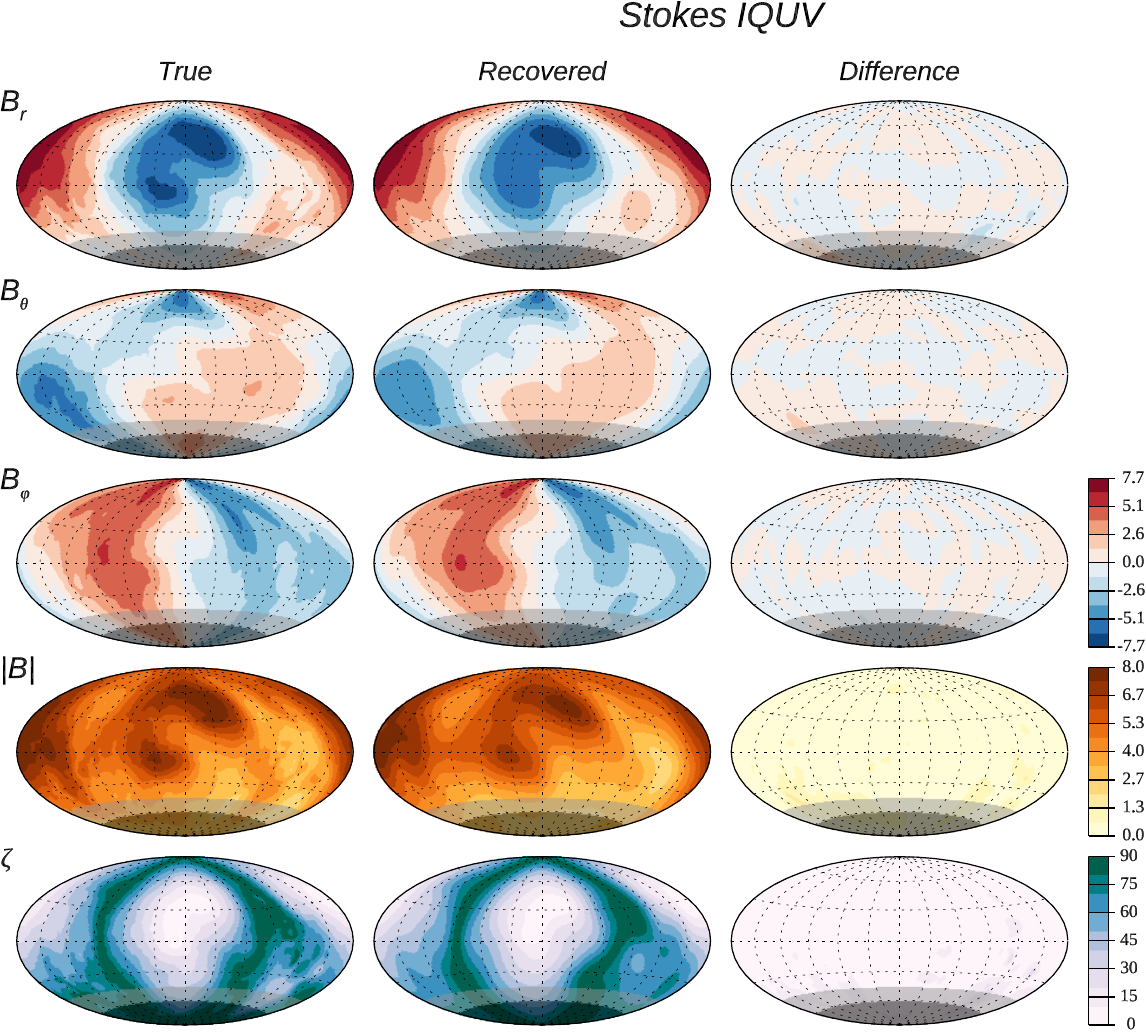}
\fihps{7.8cm}{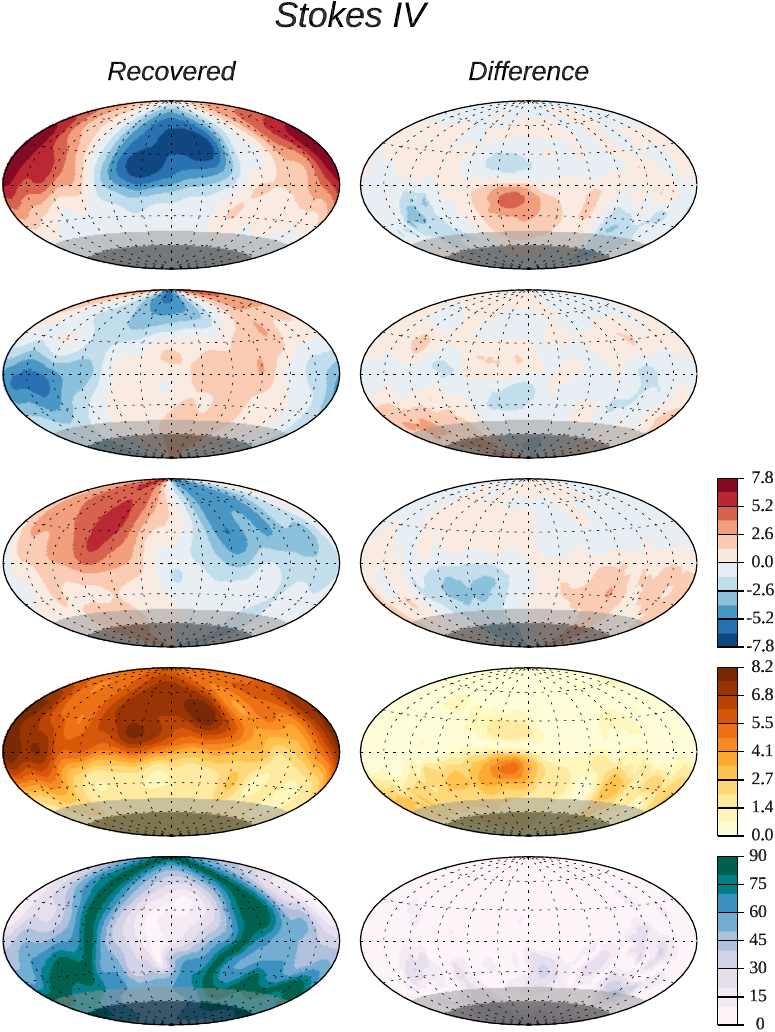}
\hspace*{0.2cm}
\fifps{3.2cm}{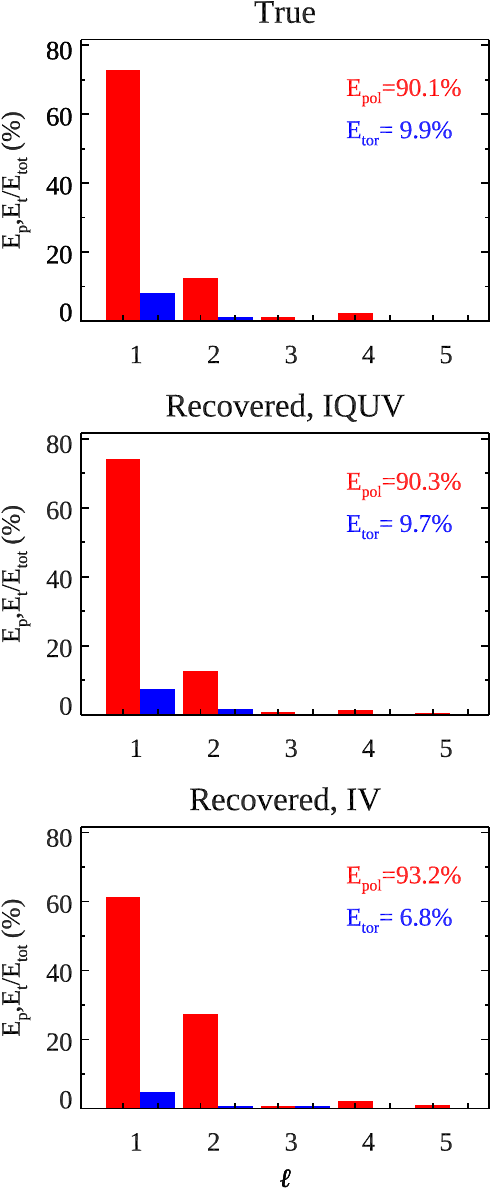}
\caption{Same as Fig.~\ref{fig:m1} but for ZDI reconstructions of model M1 using LSD profiles.}
\label{fig:m1_lsd}
\end{figure*}

\begin{figure*}[!th]
\centering
\fihps{7.8cm}{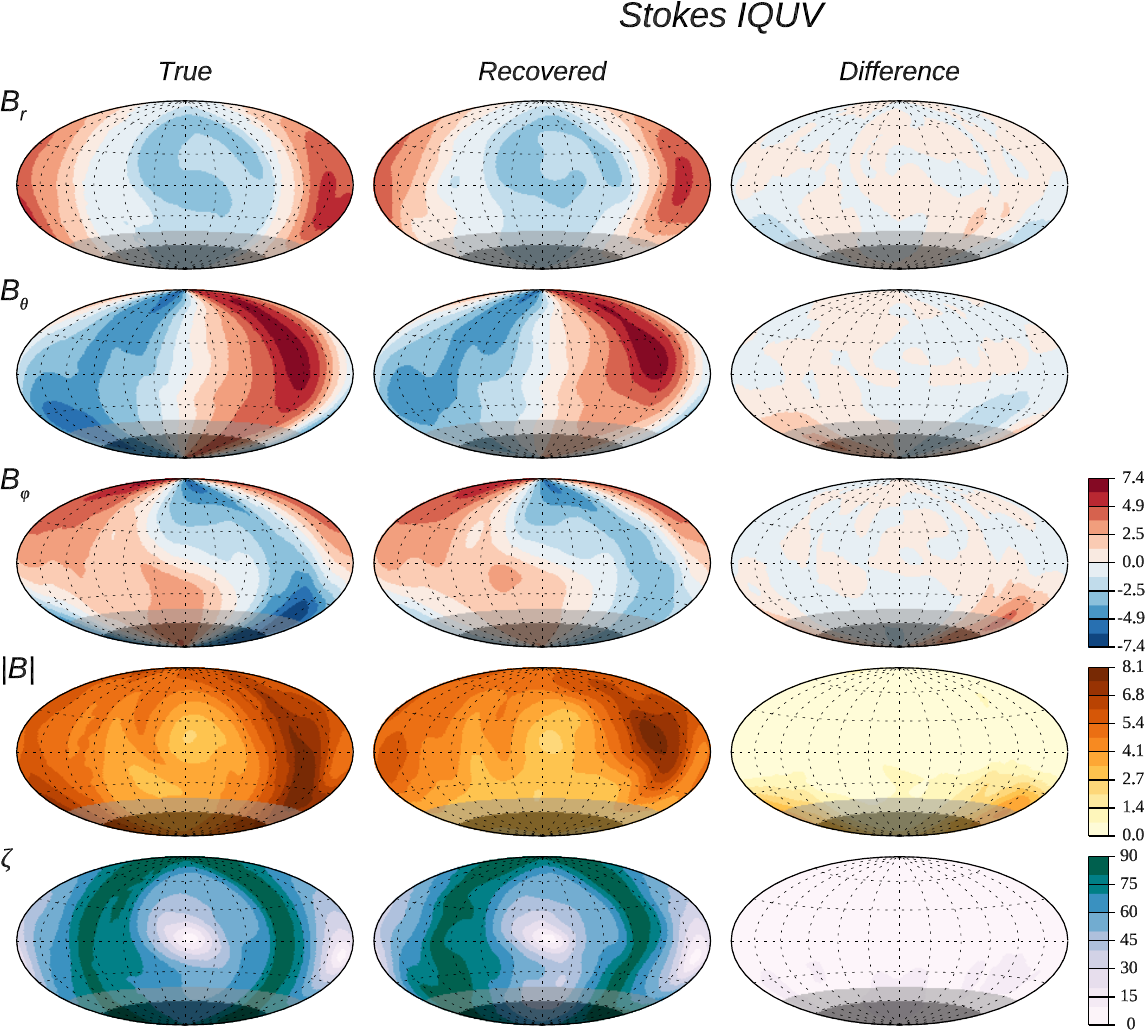}
\fihps{7.8cm}{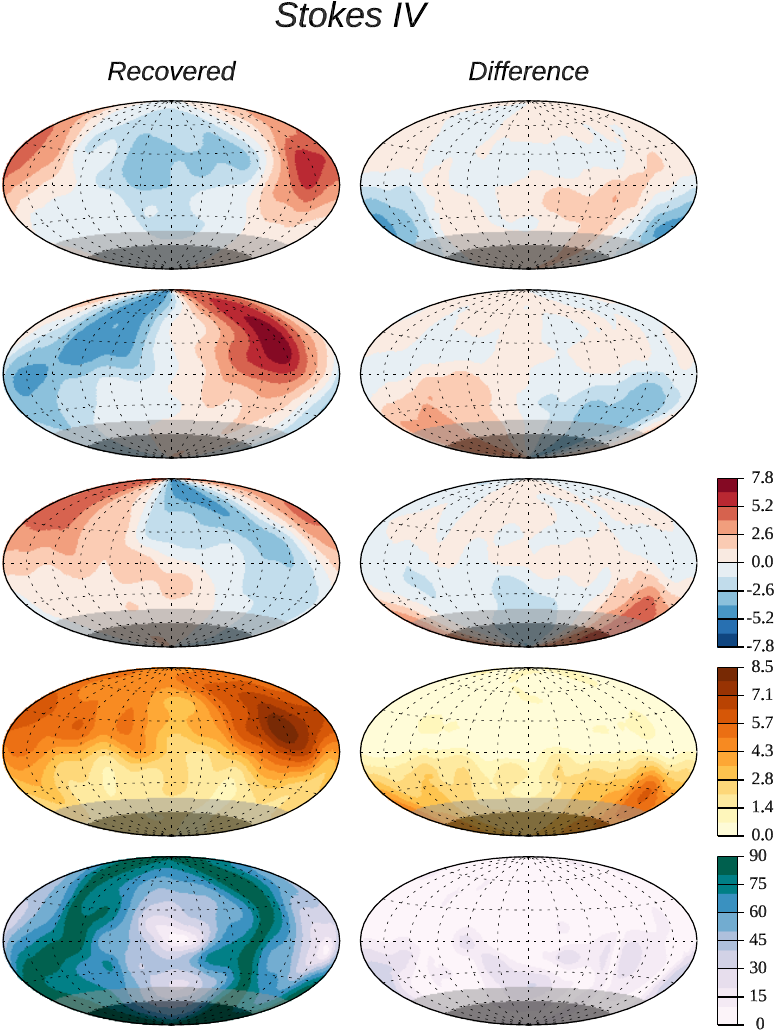}
\hspace*{0.2cm}
\fifps{3.2cm}{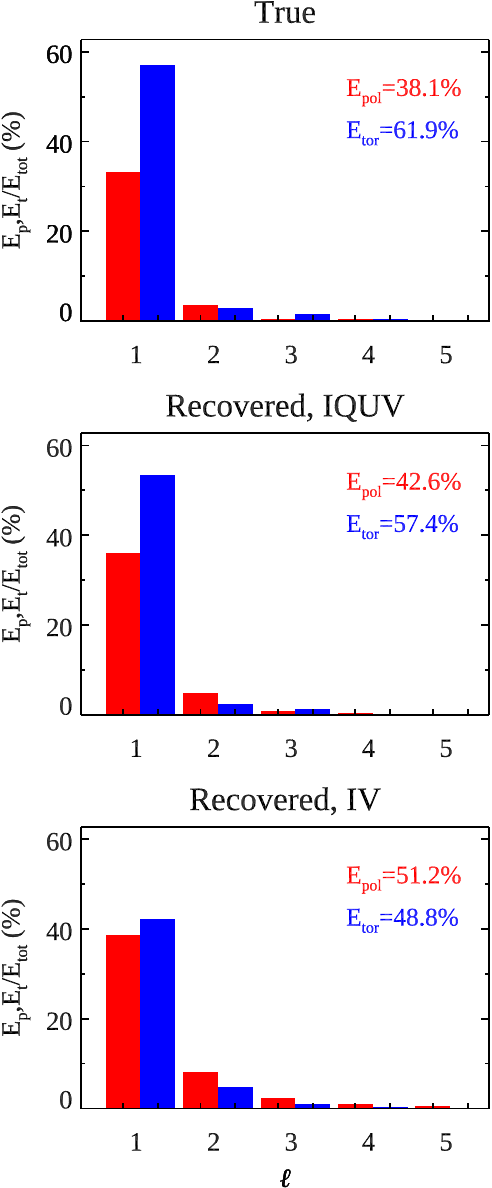}
\caption{Same as Fig.~\ref{fig:m1} but for ZDI reconstructions of model M2 using LSD profiles.}
\label{fig:m2_lsd}
\end{figure*}

\begin{figure*}[!th]
\centering
\fihps{7.8cm}{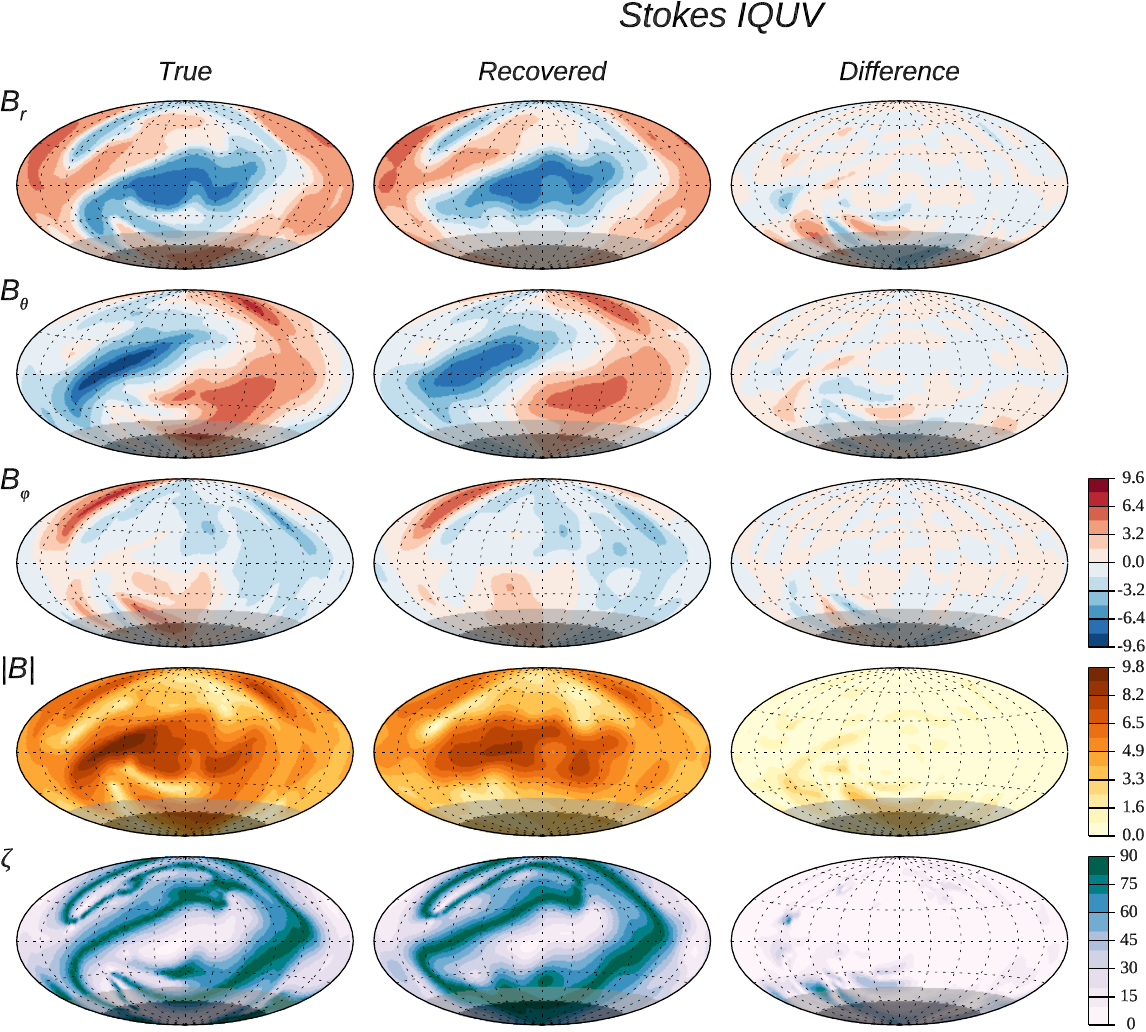}
\fihps{7.8cm}{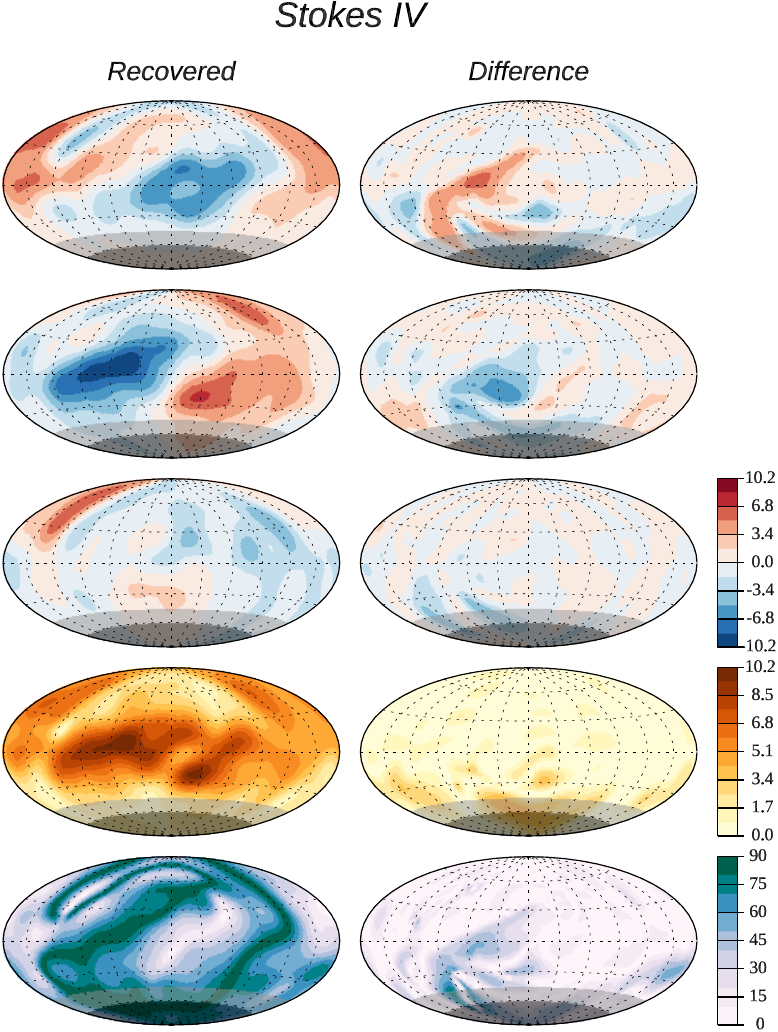}
\hspace*{0.2cm}
\fifps{3.2cm}{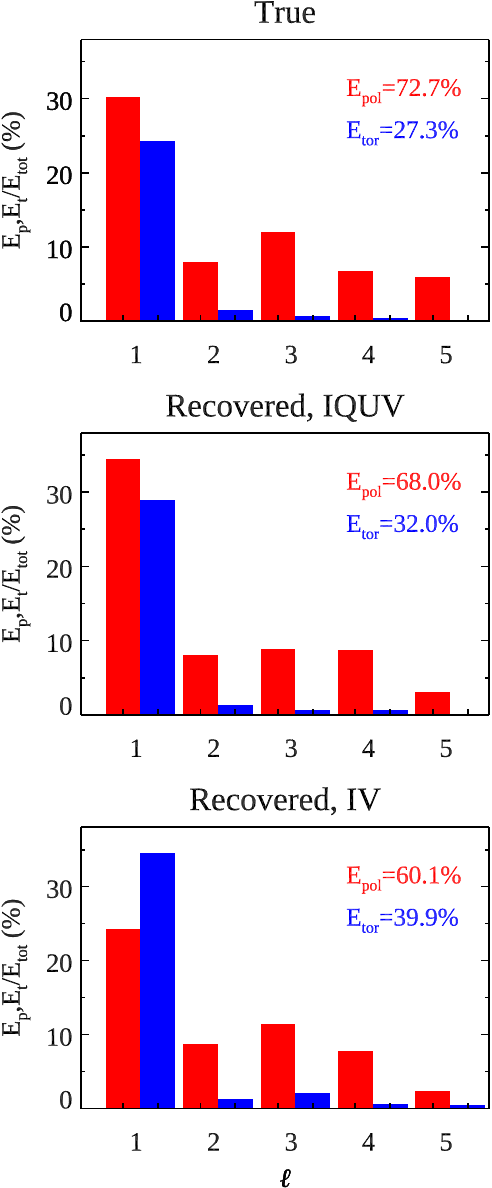}
\caption{Same as Fig.~\ref{fig:m1} but for ZDI reconstructions of model M3 using LSD profiles.}
\label{fig:m3_lsd}
\end{figure*}

\clearpage

\begin{figure*}[!t]
\centering
\fifps{4.5cm}{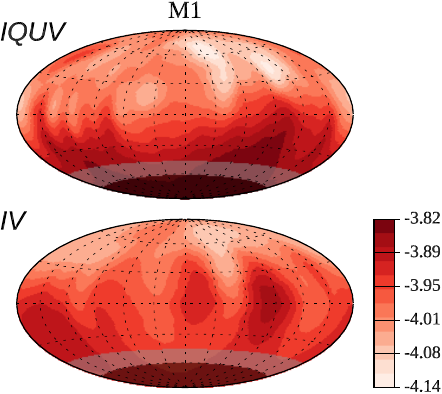}\hspace*{3mm}
\fifps{4.5cm}{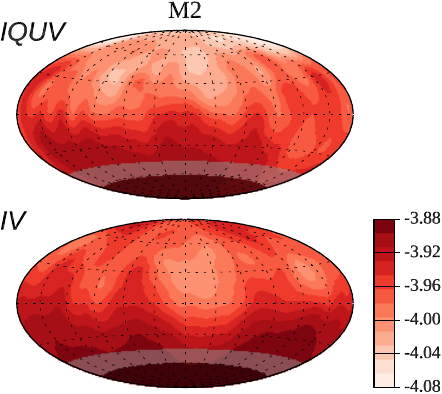}\hspace*{3mm}
\fifps{4.5cm}{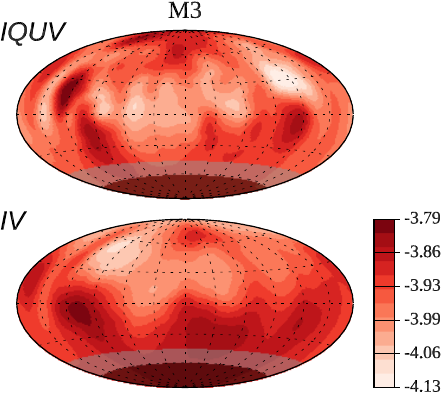}
\caption{Spurious Fe abundance inhomogeneities obtained in the Stokes $IQUV$ (\textit{top}) and Stokes $IV$ (\textit{bottom}) inversions using individual lines. In both cases, we show inversion results based on the $\alpha\ne\beta$ spherical harmonic parameterisation. The colour bars give the element abundance in $\log(N_{\rm Fe}/N_{\rm tot})$ units.}
\label{fig:abn}
\end{figure*}

\begin{figure*}[!t]
\centering
\fifps{4.5cm}{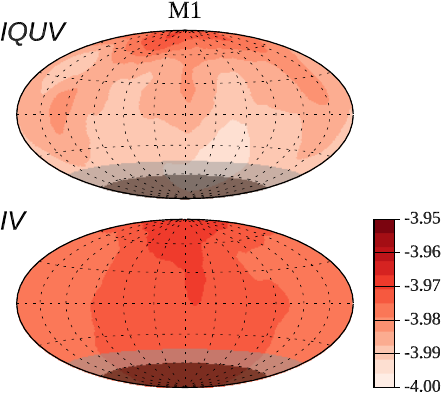}\hspace*{3mm}
\fifps{4.5cm}{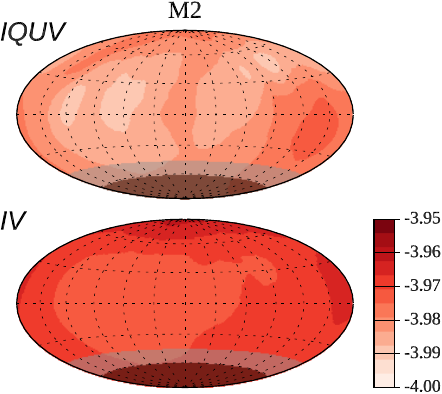}\hspace*{3mm}
\fifps{4.5cm}{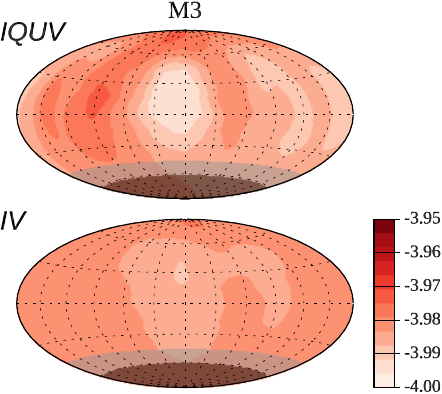}
\caption{Same as Fig.~\ref{fig:abn} but for the inversions using LSD profiles.}
\label{fig:abn_lsd}
\end{figure*}

\end{appendix}

\end{document}